\documentclass[aip,amsmath,amssymb,reprint]{revtex4-1}
\usepackage{hyperref}
\usepackage{graphicx}
\usepackage{dcolumn}
\usepackage{bm}
\usepackage[table]{xcolor}

\begin{document}
\author{Shikha Bhadoria}
\author{Naveen Kumar} \email{naveen.kumar@mpi-hd.mpg.de}
\author{Christoph H. Keitel} 
\affiliation{Max-Planck-Institut f\"ur Kernphysik, Saupfercheckweg 1, 69117 Heidelberg, Germany}

\title{Collisionless shocks in laboratory astrophysics experiments}
 
\date{\today}

\begin{abstract}
Influence of the plasma collisions on the laser-driven collisionless shock formation and subsequent ion acceleration is studied on the basis of two different collisional algorithms and their implementations in two well-known particle-in-cell codes EPOCH and SMILEI. In this setup, an ultra-intense incident laser pulse generates hot-electrons in a thick target, launching an electrostatic shock at the laser-plasma interface while also pushing the interface through the hole-boring effect. We observe, to varying degrees, the weakening of the space-charge effects due to collisions and improvements ($\ge 10\%$) in the energy spectra of quasi-monoenergetic ions in both PIC codes EPOCH and SMILEI. These results establish the `collisionlessness' of the collisionless shocks in laboratory astrophysics experiments.
\end{abstract}

\maketitle

\section{Introduction}
Collisionless shocks naturally occur in astrophysical environments such as supernova remnants, gamma ray bursts etc. \cite{Blandford:1987aa}. They can accelerate particles to very high energies and are believed to be responsible for high energy cosmic rays and non-thermal particles in astrophysical scenarios~\cite{Blandford:1987aa,Spitkovsky:2008aa,*Spitkovsky:2007aa}. Collisionless shocks can also be generated in a laboratory ~\cite{Romagnani:2008aa}.  Since collisionless shocks are efficient accelerators of particles (both leptons and hadrons), they can be used to accelerate ions in a laboratory \cite{Silva:2004aa,Haberberger:2012aa,Fiuza:2012ab,*Fiuza:2013aa} which can be beneficial for medical science, particularly for the treatment of cancer \cite{Salamin:2008aa,*Linz:2007aa}. In general, the laser produced shock waves can have a wide range of applications from nuclear fusion to material sciences \cite{Turrell:2015aa}. 

Despite their ubiquity in nature, the microphysics involved in the formation of collisionless shocks is not yet fully understood. In recent years, there have been growing number of efforts to understand the roles of various plasma instabilities in the formation of collisionless shocks in a laboratory, constituting a crucial aspect of the laboratory astrophysics research~\cite{Forslund:1970aa,*Kato:2010aa,Stockem:2014ab,*Bret:2013aa,Ryutov:2014aa,Ross:2017aa,Huntington:2015aa,Park:2015aa,Ruyer:2015aa,Fiuza:2012aa}. Such studies are subject to the scaling laws with regard to the interpretation of astrophysical observations~\cite{Ross:2017aa} but nonetheless are also complementary to the rich literature of the plasma instabilities~\cite{Weibel:1959aa,Karmakar:2009aa,Bret:2010aa}. A straightforward configuration to study the collisionless shocks, in a laboratory, is the one where two counter-propagating unmagnetized plasma flows are allowed to collide~\cite{Huntington:2015aa,Park:2015aa,Ross:2017aa}. In this set-up, the interaction region of the two plasma flows is susceptible to numerous plasma instabilities \emph{e.g.} two-stream and Weibel/current filamentation instabilities (WI/CFI). For the relativistic plasma flows, the Weibel/filamentation instability is dominant and generates a strong magnetic field ~\cite{Stockem:2014ab,*Bret:2013aa}. The particles get scattered in this magnetic field and piled up to create a shocked region. A smaller number of particles can get accelerated from this shock by the Fermi acceleration processes, yielding an energy spectrum which is a power-law distribution \cite{Spitkovsky:2008aa,*Spitkovsky:2007aa}. Since the Weibel instability generates a stronger magnetic field these shocks are also called the Weibel mediated or electromagnetic shocks~\cite{Stockem:2014ab,*Bret:2013aa}. However, experimental investigation of such shocks has been challenging as one needs very energetic laser-systems to drive relativistic collisionless plasma flows from the overdense plasma targets \cite{Ross:2017aa,Huntington:2015aa}. 

Both electrostatic and electromagnetic shocks can also be generated when an intense laser is incident on an overdense plasma target \cite{Silva:2004aa,Fiuza:2012aa,Ruyer:2015aa,Bhadoria2019}. In this configuration, the laser ponderomotive force heats the electrons and launches an electrostatic shock  at the target surface~\cite{Silva:2004aa}. Due to the difference in the electron and ion masses, there is a longitudinal electric field present at the shock-front which can reflect the background ions leading to ion acceleration \cite{Haberberger:2012aa,Fiuza:2012ab,*Fiuza:2013aa,Ruyer:2015aa,Bhadoria2019}. The hot electrons then propagate inside the target and while traversing through the target they excite a return plasma current. These counter-propagating electronic currents get filamented due to the Weibel instability and a strong magnetic field is generated. Due to this strong magnetic field generation, the electrostatic shock evolves and enters the electromagnetic phase of the shock formation~\cite{Stockem:2014ab,*Bret:2013aa,Ruyer:2015aa}. If the incident laser intensity is high $(I_0>10^{18}\,\textrm{W/cm}^{2})$, there is further compression of the plasma density at the interaction surface and the target surface also moves because of the hole-boring effect. Considerations of the plasma density compression at the interaction surface and the fact that the return plasma current is dense and has a low velocity, necessitate including collisions in the plasma dynamics. Although the shocks in the astrophysical scenarios are collisionless, the same can not be always true for the shocks generated in the laboratory astrophysics experiments. This calls for investigating the influence of collisions on the collisionless shock formation and consequently the ion acceleration in laser-plasma interaction.

\textcolor{black}{To ascertain the `collisionlessness' of the laser-driven collisionless shocks, we comprehensively study the influence of both electron-electron and electron-ion collisions on the shock formation and ion acceleration by employing the current versions of two well-known PIC codes viz. EPOCH (version 4.16.6) \cite{Arber:2015aa} and SMILEI (version 4.1)~\cite{SMILEI-v4.1}. EPOCH contains two different collision algorithms, \emph{e.g.} Sentoku-Kemp (SK)~\cite{Sentoku:2008aa} and Nanbu-Perez (NP)~\cite{Nanbu1997,Perez2012}, while SMILEI employs only the NP algorithm. Comparisons of the  EPOCH and SMILEI results can give an incontrovertible evidence of the role of collisions on collisionless shock acceleration (CSA) in laboratory astrophysics experiments.} In order to systematically quantify the impact of the collisions in both algorithms, we start with the growth of field and particle energies in the PIC simulations corresponding to the different algorithms. Then we focus on the physics of the shock formation and study the shock density jump and associated collisional weakening of the space-charge field due to collisions. Afterwards, we examine the hot-electron transportation as it is connected with the ion acceleration. We show that both SK and NP algorithms yield a higher shock density jump compared to the respective collisionless cases. The higher shock density jump is attributed to collisional weakening of the space-charge effects at the laser-plasma interface. The implementation of the SK algorithm in EPOCH (EP-SK case) somewhat causes a stronger resistive field generation and consequently a smaller penetration depth of the hot-electrons. This yields a significant improvement in the ion energy spectra. The implementation of the NP algorithm, both in EPOCH (EP-NP case) and SMILEI (SM-NP case) codes, does not generate as stronger resistive field as in the EP-SK case, at the laser-plasma interface. Since the hot-electron circulation has a bearing on the quality of shock accelerated ions, the ion-energy spectra differs in the implementation of both algorithms in PIC codes. The paper is organised as follows. \textcolor{black}{In Sec.~\ref{pic_sim} we first state the PIC simulation parameters for three different simulation runs viz. $(i)$ EP-SK case $(ii)$ EP-NP case and  $(iii)$ SM-NP case and briefly discuss the differences in the SK and NP algorithms and their implementation in PIC codes. In Sec.~\ref{dens_jump}, and \ref{pha_spa}, we discuss the impact of collisions on the shock density jump, evolution of the  electromagnetic energy densities and phase space of electrons and ions, in all three simulations runs. The impact on the ion energy spectra is finally discussed in Sec.~\ref{ion_energy} where spectra from both planar and tailored targets are included. Finally, the summary and conclusions are presented in Sec.~\ref{Conclusions}.}

\section{2D PIC Simulations}\label{pic_sim}

\begin{table}[b]
\caption{\label{table_sim}%
The simulation parameters for three simulation runs using PIC codes EPOCH and SMILEI.
}%
\begin{ruledtabular}
\begin{tabular}{lp{2cm}|p{2.0cm}|p{2.0cm}|p{1.8cm}r}
&Simulation parameters & EPOCH-4.16.6 (EP-SK)& EPOCH-4.16.6 (EP-NP) & SMILEI-v4.1 (SM-NP) & \\  \hline
&Collisional algorithms & Sentoku and Kemp (SK) & Nanbu and Perez (NP) & Nanbu and Perez (NP) \\ \hline
&Particles per cell & 50     & 50     & 49  \\ \hline
&$L_x\times L_y \,[\mu m^2] $& $120 \times 8$& $120 \times 8$& $120 \times 8$\\ \hline
& $\Delta_x\times \Delta_y$ [nm$^2$]& $ 20\times10 $ & $ 20\times10 $& $ 20\times10 $\\ \hline
&$n_e[n_c],T_{e,i}[eV]$& 50, 850  & 50, 850  & 50, 850 \\ \hline
&$T_{sim}$[fs] & 1500  & 1500  & 1300 \\ \hline
&$dt$[fs] & 33  & 33  & 33.33 \\ \hline
&$a_0,\lambda_0[\mu m]$& 60, 1 & 60, 1  & 60, 1 \\ \hline
&$I_0$[W/cm$^2$]& $5 \times 10 ^{21}$  & $5 \times 10 ^{21}$  & $5 \times 10 ^{21}$ \\ 

\end{tabular}
\end{ruledtabular}

\end{table} 

We consider a scenario where an electron-proton plasma target of density $ n_e = 50\, n_c $, is irradiated with a plane linearly polarized laser with normalized vector potential $ a_0=e E_0 / m_e \omega_0 c =60 $ and an infinite duration. Here, $E_0$ is the electric field of the laser, $ m_i/m_e=1836,$ $Z_i=1 $, where $m_e$ and $m_i$ are the masses of electron and ion respectively, $Z_i$ is the atomic number, and the critical density for a laser pulse is $n_c = m_e \omega_0^2 / 4 \pi e^2$, where $e, \omega_0$ and $c$ are the electronic charge, the laser frequency and the velocity of the light in vacuum respectively~\cite{Fiuza:2012aa,Ruyer:2015aa}.  \textcolor{black}{For the purpose of a thorough discussion on the impact two collisional algorithms on the CSA of ions, we first choose a planar target with a constant density and a thickness, $L_d= 50 \mu$m. We also employ a tailored target to show the improvement in the ion acceleration energy and spectra quality~\cite{Fiuza:2013aa,Fiuza:2012ab}.  In this case, the plasma has a maximum electronic density, $n_e(x) =  50n_c\, S(x)$, where $S(x)= x/x_1,\, x \leq x_1\,\textrm{and}\, S(x)=e^{-  (x-x_1) /l_{g}}, x >x_1$. Here $x_1 = 6\,\mu$m, upto which the density increases linearly and then decays exponentially with a scale length of $\sim 21$ microns ($l_g \sim \sqrt{m_i/m_e}\lambda_0/2$ as stipulated by Fiuza \emph{et al.}~\cite{Fiuza:2013aa}}).  The other simulations parameters are listed in Table \ref{table_sim}. The combined initial collision frequency corresponding to the simulation parameters is ($\nu_{ee} + \nu_{ei} \sim 0.01 \omega_{pe}$)~\cite{Huba2013}, where $ \omega_{pe}=(4\pi n_e e^2/ m_e)^{1/2} $ is the electronic plasma frequency. However, the collision frequencies are dynamically evaluated by the PIC code at each time-step of the simulation.

\subsection{SK and NP algorithms for plasma collisions}\label{coll_algo}
The two algorithms differ in both scattering calculation and numerical pairing of arbitrary weighted macro-particles in PIC codes. The NP algorithm calculates the scattering angle based on the cumulative scattering theory where several small angle collisions are grouped into a single binary collision with a larger scattering angle. The SK algorithm, on the other hand, randomly chooses a finite scattering angle from a Gaussian distribution with a certain variance that ensures the collision term is same as that of the Fokker-Planck equation. Also, scattering calculation in the NP algorithm is performed in only one frame of reference while the SK algorithm changes from centre-of-mass frame to one-particle-rest frame. The numerical implementation of statistics in the NP algorithm is based on the so-called `rejection method' where the macro-particle of larger weight is not always scattered, while the SK algorithm uses a `merging method' where the heavier macro-particle is always scattered undergoing only a partial scattering based on its scattering probability. \textcolor{black}{Thus, comparing the effects of collisions by both algorithms in different codes one can gain a significant insight into the role of collisions on the collisionless shock formation and subsequent ion acceleration from shocks.}

\section{Filamentation, shock density jump $\&$ weakening of the space charge field} \label{dens_jump}
\begin{figure}
\includegraphics[height=0.39\textheight,width=0.48\textwidth]{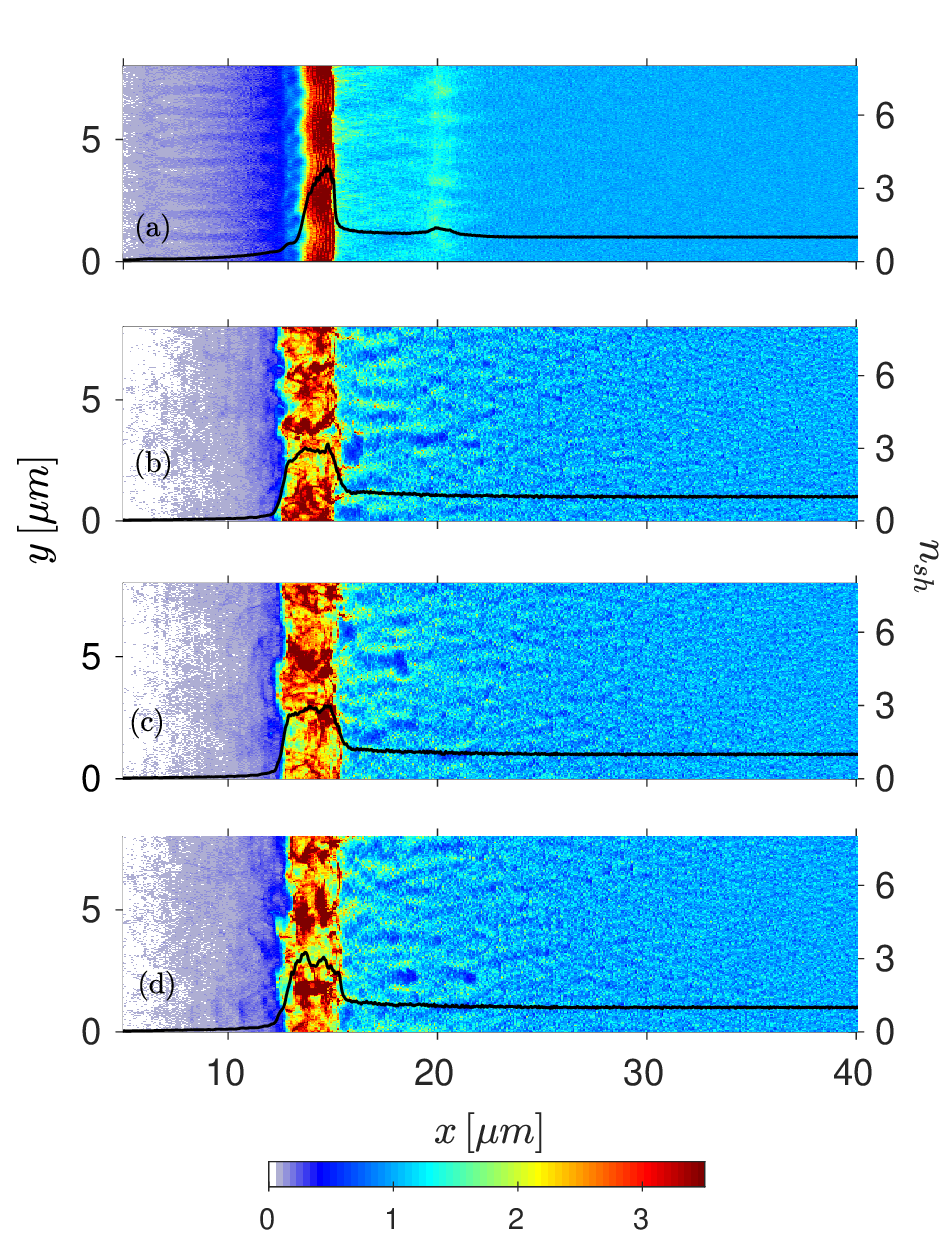}
\caption{Plasma density (normalised by initial slab density) at $198$ fs showing filamentation in all the cases. Panel (a) shows EP-SK case, (b) shows EP-NP case, (c) shows SM-NP case and (d) shows collisionless case from SMILEI. The colorbar and second $y$-axis show show the density jump associated with the shock.}
\label{num_dens2D}
\end{figure}

\subsection{Filamentation induced magnetic field generation}\label{fila}

\begin{figure}
\includegraphics[height=0.21\textheight,width=0.46\textwidth]{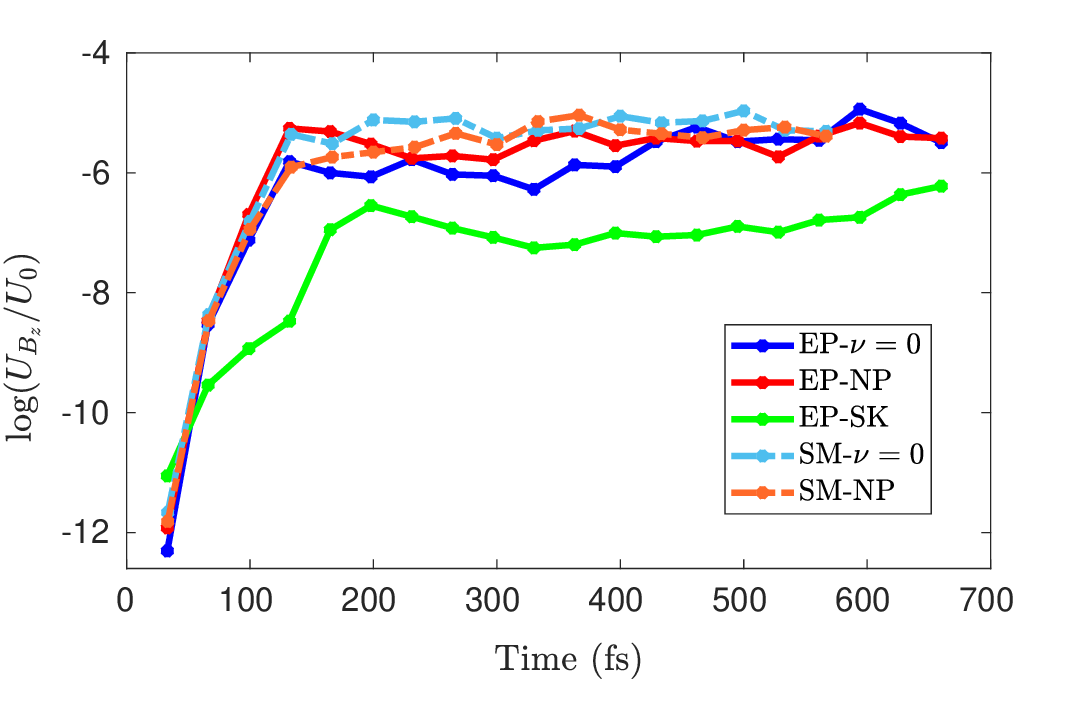}
\caption{Magnetic field energy density ($U_{B_z}$) evolution over time for collisionless EP (blue), EP-NP (red), EP-SK (green), collisionless SM (dot-dashed blue) and SM-NP (dot-dashed orange). This is normalised by laser's magnetic field energy density. }
\label{fig_eb}
\end{figure}

Fig.\ref{num_dens2D} shows plasma density evolution, depicting the shock formation at the laser-plasma interface in each case. The laser pulse hits the target from left, heating up the hot-electron in the plasma skin-depth and launches a shock with a density jump $n_{sh}=n_d/n_u\ge 3$, in each case, where $n_d$ and $n_u$ are the plasma fluid-densities in downstream and upstream regions of the shock respectively. These hot electrons carry a large current, and their propagation in a plasma is possible only if a return plasma current is excited. This return plasma current gets filamented due to the Weibel instability in three runs as seen in Fig.\ref{num_dens2D}. Since, collisions lower the growth of the WI/CFI~\cite{Karmakar:2009aa,Bret:2010aa}, thus one can expect a subdued filamentation in three cases compared to the collisionless case. Indeed comparing panels (a), (b) and (c) with panel (d) in Fig.\ref{num_dens2D}, one can see that the filamentation is strongly reduced in the EP-SK case [panel (a)] while the latter two cases viz. EP-NP and SM-NP [panels [b] and [c], respectively], the reduction in the filamentation is small compared with the respective collisionless cases [panel (d)]. It is further captured in Fig.\ref{fig_eb} where the magnetic field energy ($U_{Bz}$) associated with filamentation  is plotted in each case with time. The energy is obtained from each simulation run as $U_{B_z} = \int_0^{L_y}\int_{x_p}^{L_x}[B_z^2(x,y)/4\pi] dxdy$, where $x_p$ is the position of the laser piston which is updated after every instant with the piston velocity $v_p$ to segregate the laser magnetic field from the magnetic field generated due to the WI/CFI. $U_{B_z}$ is normalised by laser's field energy density per unit length $U_0=[B_0^2/4\pi] L_xL_y$ and is plotted on a logarithmic scale.  One can clearly notice that both the EP-NP and SM-NP cases show a smaller difference with the respective collisionless cases during the linear stage of the Weibel/filamentation instability dynamics. While the EP-SK case shows a significant reduction in the magnetic field energy development. This suggests that merging method of the SK algorithm in EPOCH code causes stronger thermalisation and results in higher effective collision frequency. Table~\ref{table_growth_rate} compares the growth rates of the WI/CFI from PIC simulations and theory (see Sec.\ref{Appendix} for details). One can immediately notice that the theoretical growth rates are a bit smaller than the PIC simulation results. However, accounting for collisions, one can see a good agreement between the theoretical and PIC values for the NP algorithm. While the implementation of the SK in EPOCH (EP-SK) case shows a larger reduction, in sync with the observations in Figs.\ref{num_dens2D} and \ref{fig_eb}. Stronger reduction in the growth rate in the EP-SK case \textcolor{black}{($\sim 30\%$)} further confirms that the implementation of the  SK algorithm in EPOCH seems to yield higher effective collision frequency \textcolor{black}{(only $\sim 10\%$ reduction in NP algorithm cases)}.
One may also note that the EP-NP case also shows a higher build up of the magnetic field energy (solid red line) compared to the respective collisionless case (solid dark-blue line) closer to the nonlinear stage ($\sim 150$ fs) of the WI/CFI, while the SM-NP case does not show this behaviour. Although the energy in the SM-NP case (dashed orange line) becomes marginally higher than the respective collisionless case (dashed light-blue line) in the nonlinear stage of the instability ($400$-$500$ fs). The strong reduction in the EP-SK case can also be attributed to the stronger weakening of the space charge field and the resistive field generation at the laser-plasma interface and is further discussed in Sec.\ref{n_sh}.
\noindent

\begin{table}[b]
\caption{\label{table_growth_rate}%
The growth rate of magnetic field energy density for all simulations (computed in the linear region of instability development \emph{i.e.}  between  $t=(33$-$66)$ fs from Fig.~\ref{fig_eb} using linear curve fitting with 95$\%$ confidence bounds). \textcolor{black}{The theoretical estimates are from the linear kinetic theory employing fitted Maxwell-J{\"u}ttner distribution with hot-electron ($n_{he}$) to return plasma current ($n_{rc}$) density ratio of $n_{he}/n_{rc}=0.3$ (See Sec.\ref{Appendix} for details)}. All growth rates are normalised by the electron plasma frequency.
}
\begin{ruledtabular}
\begin{tabular}{lp{1.8cm}|p{1cm}|p{1cm}p{1.2cm}|p{1.1cm}}
& Cases &$\delta_{max}^{kin}$ (Theo.) & \multicolumn{3}{c}{$\delta_{max}^{kin}$(Sim.)}  \\  \hline
&  & & \multicolumn{2}{c|}{EP}& SM \\
& $\nu = 0$ &0.00463 & \multicolumn{2}{c|}{0.00549} & 0.005028  \\ \hline
& & & EP-SK & EP-NP & SM-NP \\
& $\nu=0.01$ & 0.00420  & 0.00377 & 0.004823 & 0.004458  \\

\end{tabular}
\end{ruledtabular}

\end{table} 
 
\subsubsection{Field and particle energies build-up}\label{Fld_particle}
 \begin{figure}
\includegraphics[width=0.5\textwidth, height=.42\textheight]{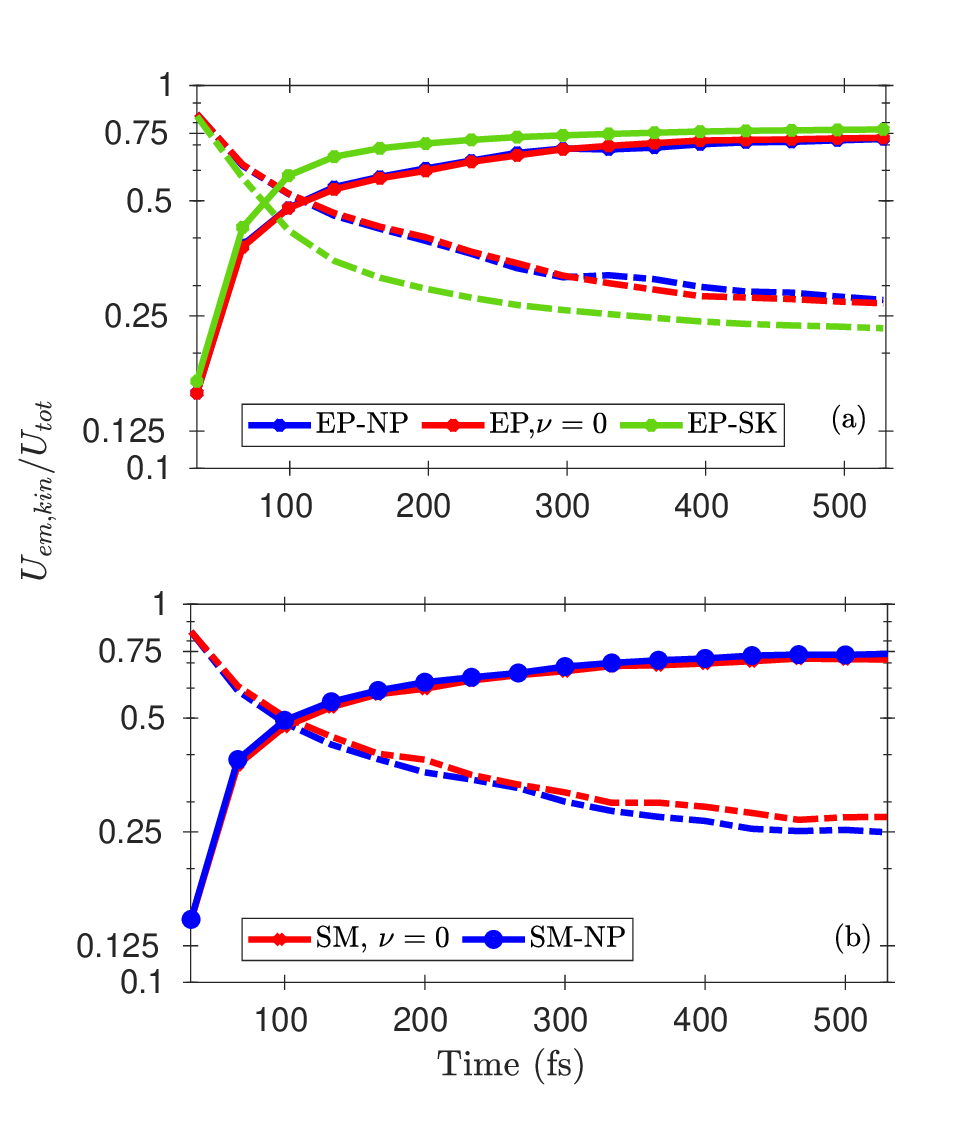}
\caption{Field energy (dash-dotted line) and particle energy (with circular markers) normalised by total energy with respect to time. Red color for the collisionless case, blue color for the NP case while green for SK case.  Panel (a) shows EPOCH code while panel (b) shows the results from SMILEI code.}
\label{Uem_kin}  
\end{figure}

To further understand the interplay of the energy equilibration between particle and field in PIC simulations in each case, we also show in 
Fig.~\ref{Uem_kin} the temporal evolution of total field energy (dashed line) and particle energy (solid line markers, both normalised by the total energy). The panel (a) shows the three run from EPOCH PIC code while the panel (b) shows the results from SMILEI code. It can be seen initially that the field energy is large (when the laser enters the simulation box) while the particle energy is low. As the laser interacts with the target, the particle energies start increasing. It begins to impart its field energy to the particles which can be seen where the field energy reduces as the particle energy increases. Eventually particle energies overtake the field energies in simulation. In EP-SK case, particles gain energy faster than the EP-NP and SM-NP cases, and consequently this equilibration time ($t_{eq}$) in smaller in EP-SK case ($t_{eq} < 100$ fs) compared to EP-NP and SM-NP cases ($t_{eq} \sim 120$ fs). Incidentally, this equilibration time  coincides with the stable shock formation in each case. Since in the EP-SK case the filamentation is subdued, particle energies (mainly of hot-electrons) are not converted into the field energy as seen in Fig.\ref{fig_eb}. Moreover, due to a stronger shock formation, as seen in Fig.\ref{num_dens2D} the energy gained by ions is also larger in EP-SK case. This is further discussed in Sec.\ref{ion_energy}.
 
\subsection{Shock-density jump and weakening of the space charge effects }\label{n_sh}
\begin{figure}
\includegraphics[height=0.18\textheight,width=0.45\textwidth]{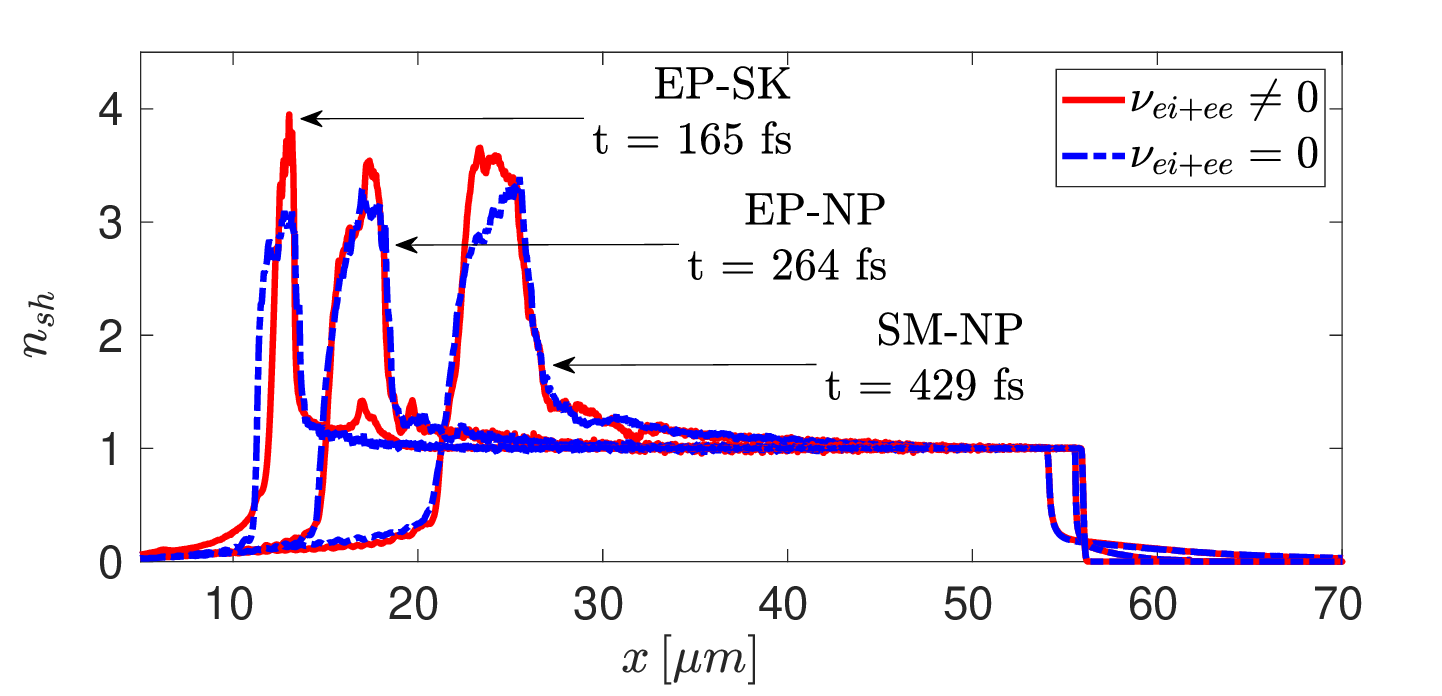}
\caption{Temporal evolution of $y$-averaged shock density jump ($n_{sh}$) for $a_0=60,\,n_e= 50n_c,\,L_d=50\mu m$ from all three simulations.}
\label{dens_jump1D}
\end{figure}

\begin{figure}
\includegraphics[height=0.37\textheight,width=0.49\textwidth]{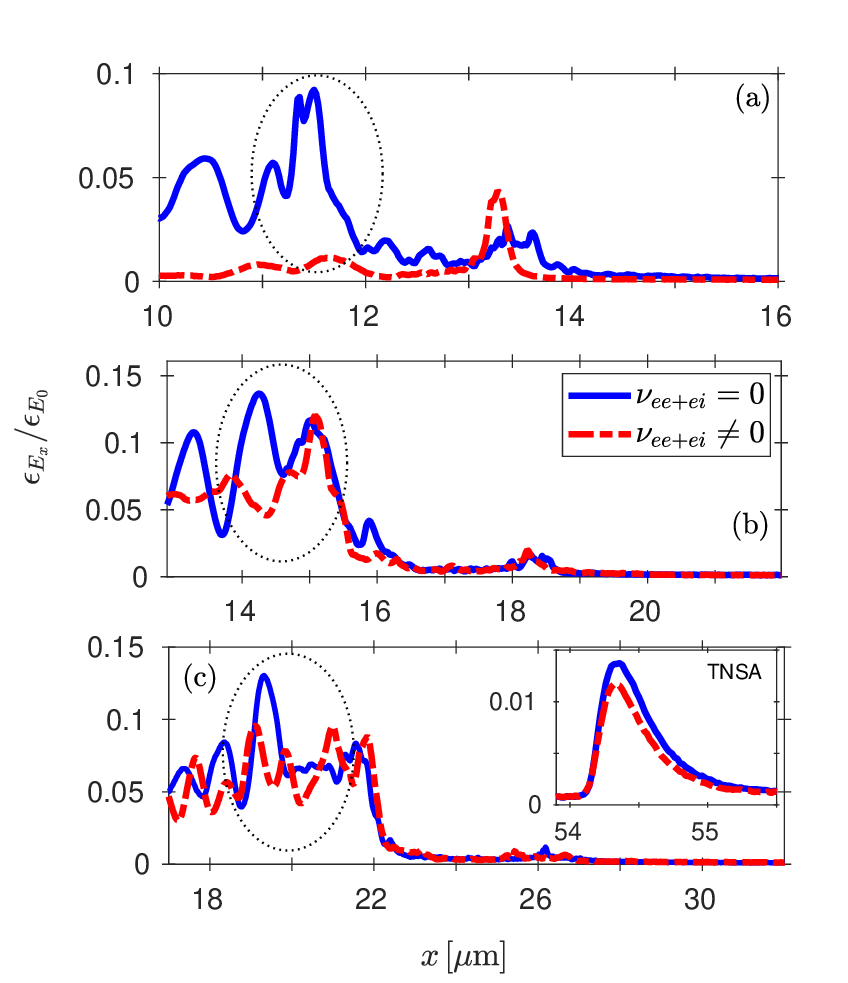}
\caption{Averaged (in $y$-direction) electric field energy (normalised by $(E_0^2/8\pi$ ) for the collisionless case (blue) and the collisional case (red) at different times.  Red line in panel (a) shows $t = 165$ fs from EP-SK algorithm, (b) shows $t = 264$ fs from EP-NP algorithm and (c) shows $t = 429$ fs from SM-NP algorithm. Each row corresponding to the three instants from three simulation runs in Fig.~\ref{dens_jump1D} . Legends are same in each case.}
\label{SCE}
\end{figure}

Fig.\ref{dens_jump1D} shows the time evolution of the transversely averaged plasma density for the case of collisionless and collisional cases for all three simulation runs, at instants when the hole-boring starts dominating. One can clearly see that the density jump is higher and the stable shock formation takes place with collisions in all the three cases. The collisional shock density jump is $n_{sh}=n_d/n_u \approx 4$, in EP-SK case. This density jump is approximately $25\%$ higher than the jump for the collisionless case ($n_{sh}  \approx 3.2$)~\cite{Blandford:1976aa,Stockem:2012aa,Ruyer:2015aa}. However, this jump is closer to the value predicted by the Rankine-Hugoniot relations for a Maxwellian plasma in high Mach numbers limit~\cite{Tidman:1971aa}. In EP-NP and SM-NP cases also, one witnesses a higher density jump compared to respective collisionless cases. Albeit, the magnitude is lower than the EP-SK case (approximately $12\%$ higher than collisionless).  Typically the density jump is calculated as~\cite{Ruyer:2015aa} $n_d/n_u=(\Gamma+1)/(\Gamma-1)$, where $\Gamma$ is the adiabatic index of the plasma. The density jump in EP-SK case can be explained by taking $\Gamma=5/3$, typical for plasma approaching thermodynamical equilibrium. While the $\Gamma=2$, corresponding to two degrees of freedom in a 2D simulation, can explain the density jump  in EP-NP and SM-NP cases. Physically, the higher density jump is attributed to the collisional weakening of the space-charge effects. The space charge field (in 1D approximation) is, $E_x=-\Delta \phi/\Delta x$, where $\phi$ is the electrostatic potential which depends on the plasma density and charge, and $\Delta x$ is the average separation between electron and ion layers. Binary collisions between two plasma particles cause scatterings of the particles, making the average spacing between two plasma particles (for example electron-ions) larger. Electron-ions collisions can enlarge $\Delta x$, and, on assuming $\phi$ to be a constant, thereby weaken the space-charge effects.  Consequently, the laser ponderomotive force can compress the plasma density to a higher value. When the hole-boring velocity acquires a constant value and no further density compression is possible, a stable electrostatic shock with a higher density jump is formed that propagates inside the plasma. Since the SK algorithm scatters the larger weight macro-particle, the electron density, upon compression at the target interaction surface, suffers larger scattering than the NP algorithm. This leads to stronger collisional weakening of the space charge field in the EP-SK case. The weakening of the space-charge effect in each case is best captured in the  development of the energy associated with the longitudinal electric field. Fig.\ref{SCE} shows the  $y$-averaged  energies, $\langle\varepsilon_{E_x}\rangle$  associated with electric field ($E_x$) at different instants. Comparisons of the  energy  associated with the longitudinal electrostatic field ($\langle\varepsilon_{E_x}\rangle$) reveals that it is significantly lower behind the shock front (marked by the dotted ellipse) at all times and in both SK and NP collisional algorithms. This is due to the collisional weakening of space charge field as discussed before. This results in a higher density compression by the laser ponderomotive force, causing  a higher shock density jump  at $t=165,264,429$ fs as seen in Fig.\ref{dens_jump1D}. In Fig.\ref{SCE}(a) ($t=165$ fs), the longitudinal electric field energy,  $\langle\varepsilon_{E_x}\rangle$ at the shock front is higher than the collisional case. One may note that the weakening of a space charge effect is a dynamic effect.
\noindent

\begin{figure}
\includegraphics[height=0.37\textheight,width=0.49\textwidth]{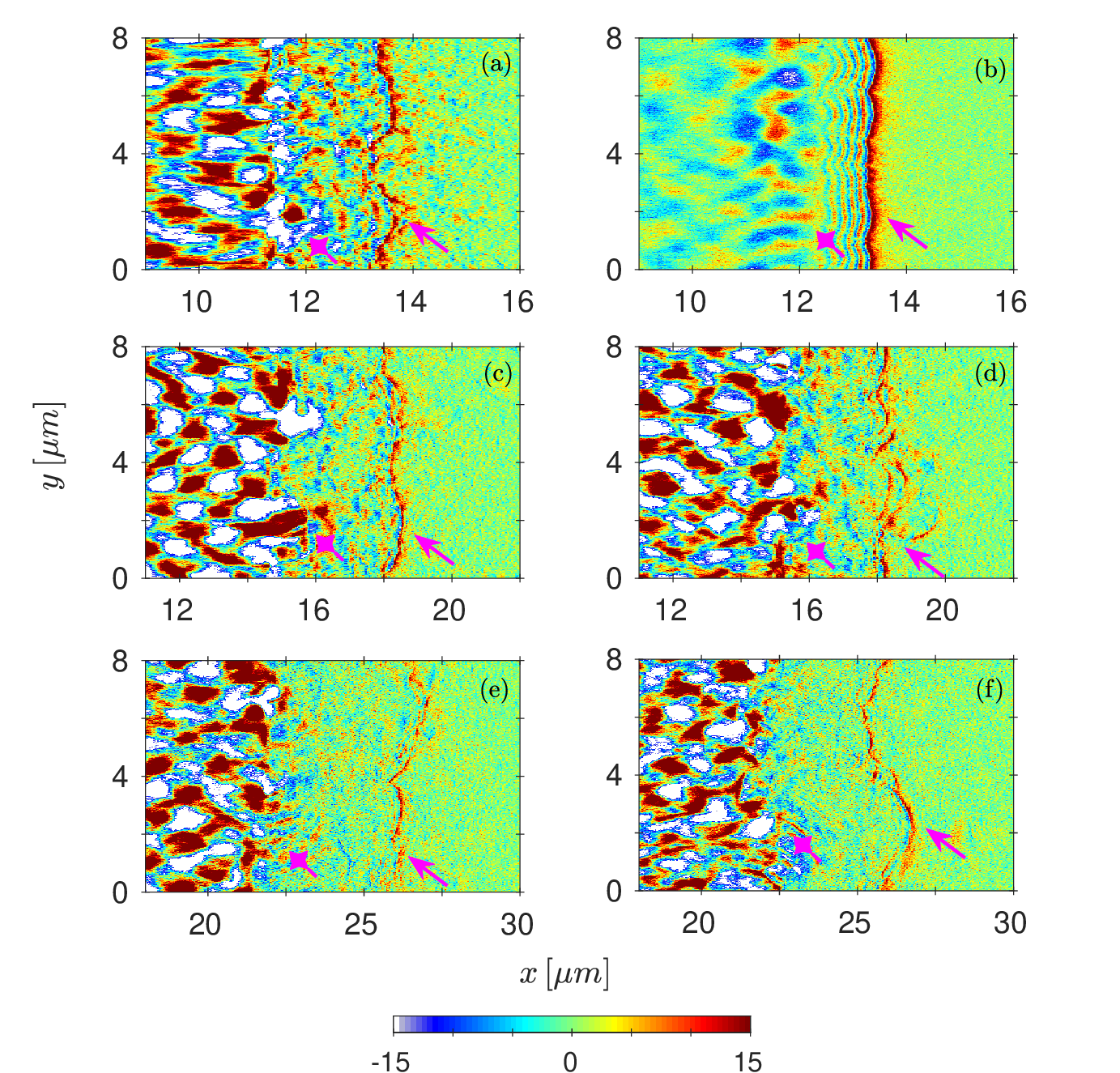}
\caption{\textcolor{black}{Electric field (normalised by $m_e c\omega_0/e$ ) for the collisionless case [first column, panels (a,c,e)] and the collisional case [second column, panels (b,d,f)] at different times from both codes and algorithms.  Panel (a) shows the collisionless case at t = 165 fs from EPOCH, panel (b) shows  collisional case with collisions implemented by EP-SK at the same time, panel (c) shows the collisionless case at t = 264 fs from EPOCH, panel (d) shows  collisional case with EP-NP at the same time, panel (e) shows the collisionless case at t = 429 fs from SMILEI and panel (f) shows  collisional case with SM-NP at the same time. The arrows with pointed heads in each panel point at the shock front whereas the arrows with star heads point at the piston region.}}
\label{ex2d}
\end{figure}

\subsection{Resistive field generation and Ohmic stopping of the hot-electrons}\label{resistive_fields}

As mentioned before, the hot-electrons generated carry a large electric current and its transport in the plasma is only possible if the hot-electron current is neutralised by the return plasma current. In a simplest form, the generation of an inductive electric field that drives the return current can be expressed by the Ohm's law in a collisional plasma $\bm{E}_{\textrm{ohm}}=\eta \bm{J}_h$, where $\eta$ is Spitzer resistivity and $\bm{J}_h$ is the current density of the hot-electrons. In collisionless plasmas, $\eta$ can denote the effective resistivity arising due to the scattering of plasma particles with the self-generated fields. This electric field $\bm{E}_{\textrm{ohm}}$, causes Ohmic stopping of the hot-electron beam and its role in reducing the hot-electron penetration depth in EP-SK case is  further discussed later in Sec.\ref{pha_spa}. Fig.\ref{ex2d} shows the Ohmic field generation in all three cases and one can notice a stronger Ohmic field generation in EP-SK case compared to EP-NP and SM-NP cases, in sync with the observations in Fig.\ref{SCE}(b,c). Comparing collisionless [panels (a,c,e)] with the respective collisional cases [(panel (b) with EP-SK, panel (d) with EP-NP and panel (f) with SM-NP], one can affirm the observations in Fig.\ref{SCE}. The position of laser piston is marked with a star-shaped arrow whereas the position of shock front is marked with a pointed arrow in each subplot of Fig.~\ref{ex2d}. Thus, one can conclude that the implementation of the SK algorithm in EPOCH code (EP-SK case) overestimates the inductive electric fields generated by the collisional plasma. 

\section{Electrons and ion phase spaces}\label{pha_spa}

\begin{figure}
\includegraphics[height=0.33\textheight,width=0.52\textwidth]{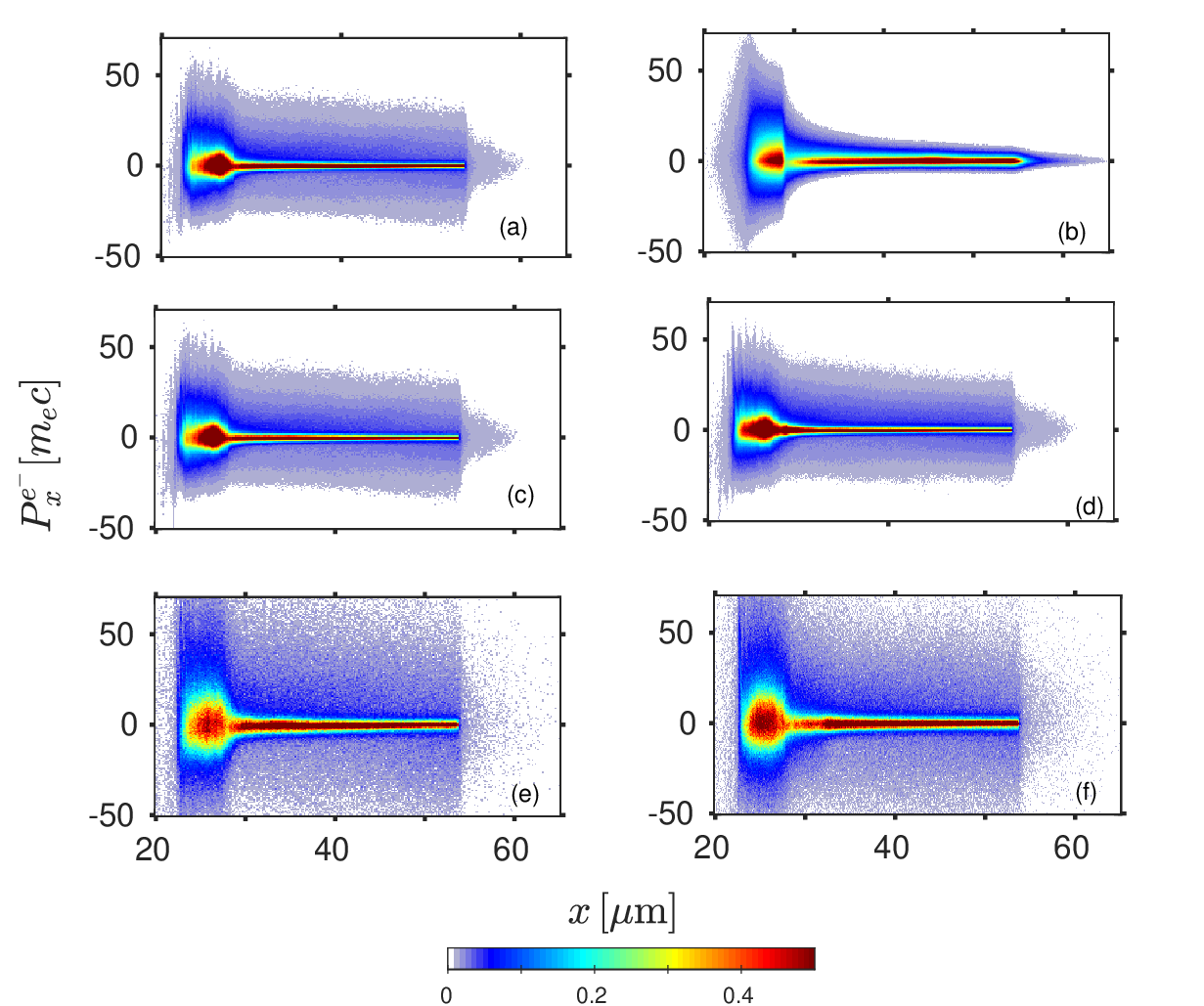}
\caption{Phase space of electrons for collisionless [first column (a,c,e))] and collisional [second column (b,d,f)] plasmas. In the second column, panel (b) is EP-SK case, panel (d) is EP-NP case and the panel (f) is SM-NP case.}
\label{xpx_elec}
\end{figure}

\begin{figure}
\includegraphics[height=0.33\textheight,width=0.52\textwidth]{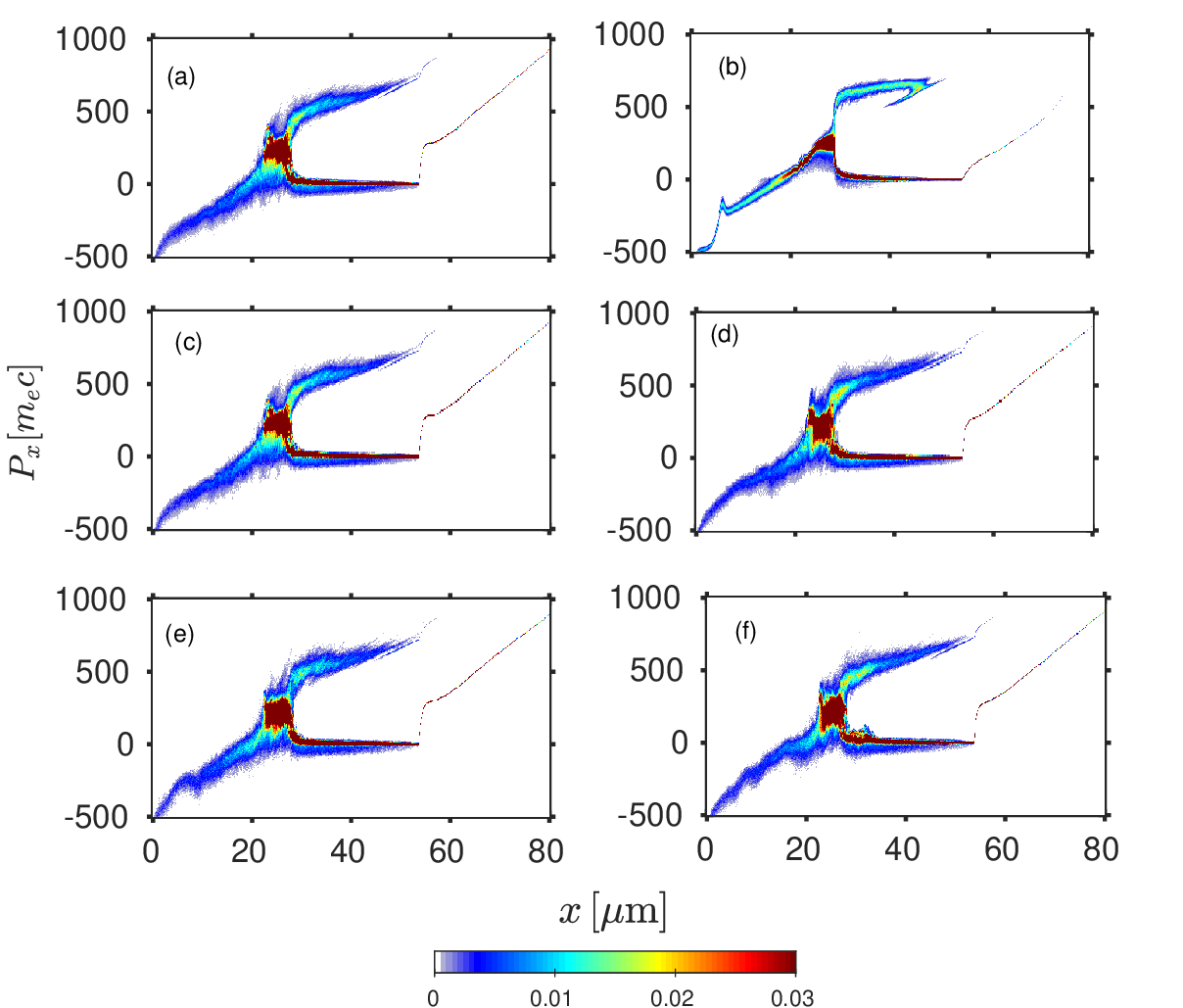}
\caption{Phase space of ions for collisionless [first column (a,c,e))] and collisional [second column (b,d,f)] plasmas at 462 fs. In the second column, panel(b) is EP-SK case, panel (d) is EP-NP case and the panel (f) is SM-NP case.}
\label{xpx_ion}
\end{figure}

In order to quantify the impact of collisions on the collisionless shock acceleration of ions, we plot in Figs.~\ref{xpx_elec} and \ref{xpx_ion}, both electron and ion phase-spaces, respectively for collisionless and collisional targets at different instants. One can clearly notice that for collisionless target (Fig~\ref{xpx_ion} first column), that there are two groups of ions being accelerated and the TNSA of ions (from the back of the target) is stronger than the collisionless shock acceleration (CSA) of ions (from the shock front). Later on these two groups of ions merge, leading to a broader ion energy spectra as observed before \cite{Fiuza:2012ab,*Fiuza:2013aa}. In the EP-SK case (first row), one sees a significant reduction of the TNSA of ions. While in the EP-NP and SM-NP cases, the TNSA of ions remains unchanged compared to the respective collisionless cases (only subtle suppression of TNSA field by collisions in SM-NP, see inset in Fig.\ref{SCE}(c)). Since TNSA of ions is connected with the hot-electron transportation in the plasma, it is instructive to examine the phase space of electrons which is shown in Fig.\ref{xpx_elec} in three cases. Propagation of the hot-electron beam (carrying large amount of current) is affected by self-generated fields of the hot-electron beam and the background plasma resistivity ~\cite{Nakatsutsumi:2018aa,Bell:1997aa,Batani:2002aa,McKenna:2013aa,*Gibbon:2005aa,*Robinson:2014aa}. It can also be affected by the magnetic field at laser-interaction surface~\cite{Nakatsutsumi:2018aa}. One can see that in the EP-SK case, the hot-electron transportation inside the target is severely affected. There is significant collimation of the hot-electron beam,  due to a strong magnetic field associated with it as seen in Fig.\ref{SCE}(a). This is in line with the results of Nakatsutsumi \emph{et al.}~\cite{Nakatsutsumi:2018aa}, where the transportation of the hot-electron flux is shown to be inhibited by the self-generated magnetic field at the interface.  Also, the hot-electron flux at the target surface shows a significant reduction. While in the  EP-NP and SM-NP cases, no such significant effect is observed, consistent with Fig.\ref{xpx_ion}. Though, compared to the EP-NP case, the angular divergence of the hot-electron beam is significantly higher in the SM-NP case, implying the influence of the self-generated magnetic field on hot-electron transportation is weaker in the SM-NP case. Nevertheless, the TNSA of ions is clearly visible in the last two rows of Fig.\ref{xpx_ion}. The result in EP-SK case can further be explained by the Ohmic stopping of the hot-electrons. Owing to finite resistivity of a collisional target, a large part of the hot-electron energy is converted into the longitudinal electric field at the shock front [Fig.\ref{SCE}(a)] required for driving the return plasma current, leading to stopping of the hot-electrons beam in a small distance. Following Bell \emph{et al.}~\cite{Bell:1997aa}, the stopping range of the hot-electrons is estimated to be $x_s \sim (250-400)\, \mu$m. On each recirculation, hot-electrons lose energy while leaving the target and can not sustain the TNSA of ions as shown in upper panel of Fig.\ref{xpx_elec}. Since one observes large deviations between the SK and NP algorithms, one can draw the conclusion that the implementation of the SK algorithm in EPOCH (EP-SK case) seems to exhibit stronger effects of the self-generated magnetic field and the Ohmic stopping compared to the implementation of the NP algorithm in EPOCH and SMILEI codes.

\section{Ion energy spectra}\label{ion_energy}

\begin{figure}
\includegraphics[width=0.51\textwidth, height=.22\textheight]{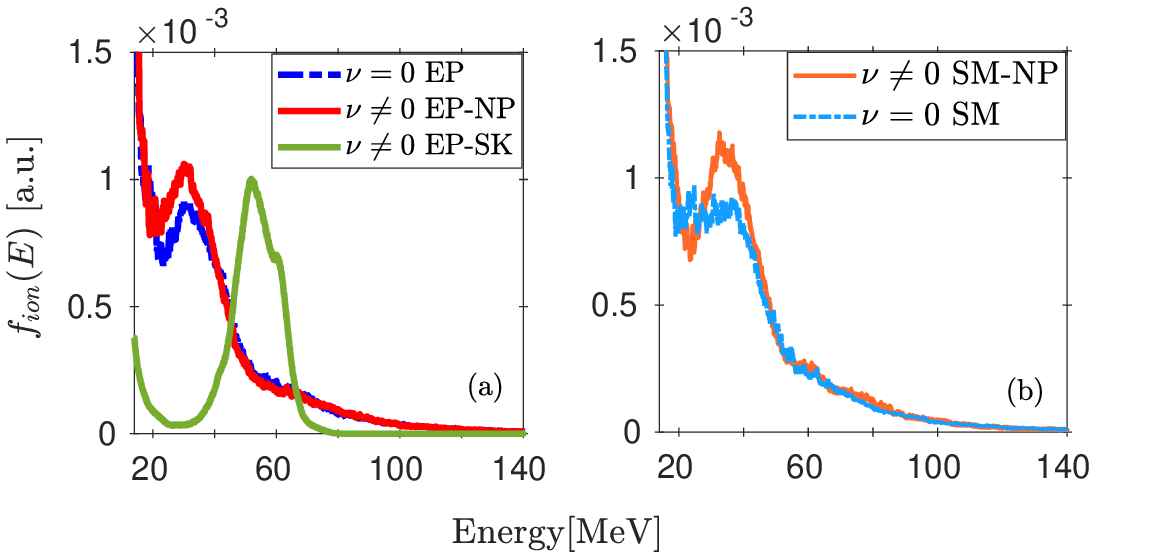}
\caption{(a) Energy spectrum of the shock-reflected ions from  collisional and collisionless plasma target at $\sim 627$ fs from EPOCH code (both EP-SK and EP-NP cases are plotted). Panel (b) shows the results from SMILEI code at the same time.}
\label{fE}  
\end{figure}
\begin{figure}
\includegraphics[width=0.5\textwidth, height=.21\textheight]{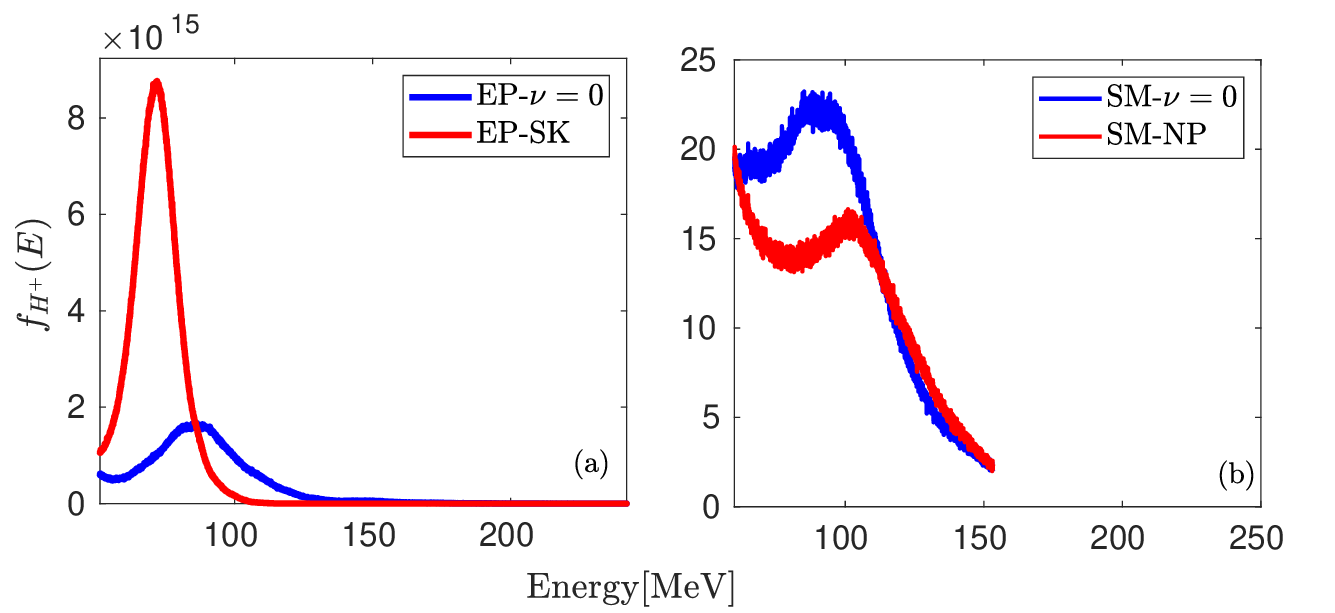}
\caption{(a) Energy spectrum of the shock-reflected ions from  collisional and collisionless `tailored' plasma target at $\sim 627$ fs from EPOCH code. Panel (b) shows the results from SMILEI code at the same time.}
\label{fig_tail}  
\end{figure}

The stronger difference between the hot-electron transport in SK and NP algorithms manifests itself in a strong difference in the ion-energy spectra as shown in Fig.~\ref{fE}(a). \textcolor{black}{In EP collisionless case, ions gain a peak energy of $\sim 31$ MeV with an energy spread $\Delta E$ of $\sim 23$ MeV whereas in SM collisionless case, ions gain a peak energy of $\sim 33$ MeV with a similar energy spread $\Delta E$ of $\sim 24$ MeV .} One can clearly see the significant improvements in the ion energy spectrum in EP-SK case \textcolor{black}{(where the energy spread becomes $\Delta E \sim 16$ MeV with higher peak energy being $\sim 53$ MeV)}, while one also observes improvements in the ion energy spectrum from NP scattering algorithm both in EP-NP \textcolor{black}{(with energy spread $\Delta E \sim 19.6 $ MeV with peak energy being $\sim 31.2$ MeV)} and SM-NP \textcolor{black}{(where the energy spread $\Delta E \sim 21.8$ MeV with peak energy being $\sim 34$ MeV)} cases. The stronger improvement in the ion-energy spectra in EP-SK case is attributed to the stronger shock formation and the Ohmic stopping of the hot-electrons. Since a stronger shock accelerates ions to same energy creating a mono-energetic ion spectra. The hot-electrons excite TNSA of ions at the back of the target. Hence partial stopping of the hot-electron flux upon recirculation in the target causes a weaker TNSA as seen in Fig.\ref{xpx_ion} (upper row). Thus, the TNSA ions do not mix up with the shock accelerated ions and the final spectrum of ions retains its quasi-mono-energetic profile. The improvements in the EP-NP and SM-NP cases can be attributed to the stronger shock generation with a slightly higher density jump. The stronger shock accelerate the ions without a significant dissipation. This results in a clearly identifiable quasi-monoenergetic ion peak compared to the collisionless targets; see Fig.\ref{fE}. One may note here that accounting for collisions leads to the quasi-monoenergetic ion acceleration in both collisional algorithms. \textcolor{black}{It may also be seen that the energy gained by the ions in the EP-SK case is higher. This could also have been expected from Fig.~\ref{Uem_kin}, where particles are seen to gain significant more energy in the EP-SK case. Although the energy spread of the ions is large, collisions have been found to improve the ion energy spectrum.  Nevertheless, the tailored targets can further optimize both the ion acceleration energies and the spectra quality.}
\textcolor{black}{Tailoring the plasma target with an exponentially decaying density profile can render the sheath field at the target's end to be low and uniform. This can further improve the ion energy gain as well as the ion beam's profile~\cite{Fiuza:2013aa,Fiuza:2012ab}. Fig.\ref{fig_tail} shows the ion energy spectrum from EPOCH code in Fig.\ref{fig_tail}(a) and SMILEI ones in  Fig.\ref{fig_tail}(b). The ions in collisionless case in Fig.\ref{fig_tail}(a) have a peak energy of about $E_{peak}\sim 86$ MeV with an energy spread $\Delta E/E \sim 35 \%$. Again collisions lead to a significant enhancement in the energy spectrum in EP-SK case with  $E_{peak}\sim 72$ MeV with an energy spread $\Delta E/E_{peak} \sim 23 \%$. Collisions also improve the ion energy spectrum in the SM-NP case. The protons from collisionless target gain $E_{peak}\sim 91$ MeV with an energy spread $\Delta E/E \sim 40 \%$ in SMILEI code. In the SM-NP case, ions gain $E_{peak}\sim 103$ MeV with an energy spread $\Delta E/E \sim 31 \%$ similar to the spreads observed recently in an experiment~\cite{Pak2018}}.

\section{Conclusions}\label{Conclusions}

To summarise, we have examined the shock acceleration of ions in a realistic scenario where the effect of the plasma collisions is indeed important. The impact of collisions is studied by employing the current versions of two different PIC codes (EPOCH and SMILEI) utilising implementations of two different scattering algorithms. Collisions influence the ion acceleration process in an indirect manner and the shock front, in the case of a mildly collisional plasma, exhibits a higher density jump than in a collisionless plasma and FWHM improvements ($\ge 10\%$) in the ion energy spectra from both PIC codes EPOCH and SMILEI. The implementation of the SK algorithm in EPOCH shows a significant enhancement in the ion-energy spectra while the small enhancements are also noticed in the NP algorithm's implementation in EPOCH and SMILEI codes. Thus, collisions do affect the collisionless shock formation and their influence  on laser-driven shock acceleration of ions, in ongoing laboratory astrophysics experiments, requires further experimental investigations.

\section{APPENDIX for calculating the growth rate of the WI/CFI}\label{Appendix}

The growth rate of the Weibel/filamentation instability is calculated using kinetic theory by employing the drifting ultra-relativistic distribution function (Maxwell-J{\"u}ttner distribution) of the form  $f(\textbf{p})=\sum_{i=he,rc} \mu_i[4\pi\gamma_i^2K_2(\mu_i/\gamma_i)]^{-1}\exp[{-\mu_i(\gamma(\textbf{p})-\beta_{he}p_x)}]$, for hot-electrons ($hc$) and return plasma current ($rc$). Here $\mu_i=m_e c^2/k_BT_i, \, \beta_i=\langle p_{x}/\gamma \rangle_j$ and $\gamma_i$ are thermal parameter, longitudinal mean drift velocity, the Lorentz factor of each $i$'th component, respectively~\cite{Bret_2010_exact}. The linearization of the relativistic Vlasov equation with an additional Krook's collision term yields the folowing dispersion relation:
\begin{equation}
(\omega^2\epsilon_{xx}-k^2)(\omega^2\epsilon_{yy})-(\omega^2 \epsilon_{yx})^2=0,
\end{equation}
where the $\epsilon_{\alpha \beta}$ are the dielectric tensor elements given by
\begin{equation}
\omega^2\epsilon_{yy}= \omega^2 + \sum_{i=he,rc} \alpha_i \frac{\omega}{k} \mathcal{B}_i,
\end{equation}
\begin{equation}
\omega^2\epsilon_{xx}= \omega^2 + \sum_{i=he,rc} \kappa_i - \delta_{i,rc} \frac{\iota\nu_0}{k}\alpha_i \beta_i \mathcal{D}_i +\alpha_i \frac{\omega}{k} \mathcal{A}_i, 
\end{equation}
and
\begin{equation}
\omega^2\epsilon_{yx}= \sum_{i=he,rc} \alpha_i \frac{\omega}{k} \upsilon_i \mathcal{C}_i.
\end{equation}
Taking $\Lambda_i=\mu_i[4\pi\gamma_i^2K_2(\mu_i/\gamma_i)]^{-1}$, in the dielectric tensors, we define $\alpha_i=2\pi n_i\mu_i\Lambda_i$, $\upsilon=\mu_i\beta_i$, $\kappa_i=n_i\mu_i\beta_i^2$, $\nu_0$ is the collision frequency and
\begin{equation}
\mathcal{A}_i=\int_{-1}^{1} du \frac{\gamma e^{-h_i}}{h_i^5} \frac{(\rho_i^2+2\upsilon_i^2) (1+h_i)+ (\upsilon_ih_i)^2}{u-\frac{\omega+\delta_{i,rc}\iota\nu_o}{k}},
\end{equation}

\begin{equation}
\mathcal{B}_i=\int_{-1}^{1} du \frac{\gamma^3 u^2 e^{-h_i}}{h_i^5} \frac{(2\rho_i^2+\upsilon_i^2 )(1+h_i)+(\rho_ih_i)^2}{u-\frac{\omega+\delta_{i,rc}\iota\nu_o}{k}},
\end{equation}

\begin{equation}
\mathcal{C}_i=\int_{-1}^{1} du \frac{\gamma^2 u\rho_i e^{-h_i}}{h_i^5} \frac{3(1+h_i)+ h_i^2}{u-\frac{\omega+\delta_{i,rc}\iota\nu_o}{k}} 
\end{equation}
and
\begin{equation}
\mathcal{D}_i=\int_{-1}^{1} du \frac{\gamma^2 \upsilon_i \rho_i e^{-h_i}}{h_i^5} \frac{3(1+h_i)+ h_i^2}{u-\frac{\omega+\delta_{i,rc}\iota\nu_o}{k}}. 
\end{equation}
Here, $\gamma=1/\sqrt{1-u^2}$, $\rho_i=\mu_i\gamma$,$\upsilon_i=\mu_i\beta_i$, $h_i=\sqrt{\rho_i^2-\upsilon_i^2}$ and $\delta$ is the Kroneckar delta (that ensures collision contribution to only be considered in the cold and dense return current). In the case of $\nu_0=0$, the above dispersion relation reduces to the one in Ref.~\cite{Bret_2010_exact}.

\begin{figure}
\includegraphics[height=0.22\textheight,width=0.48\textwidth]{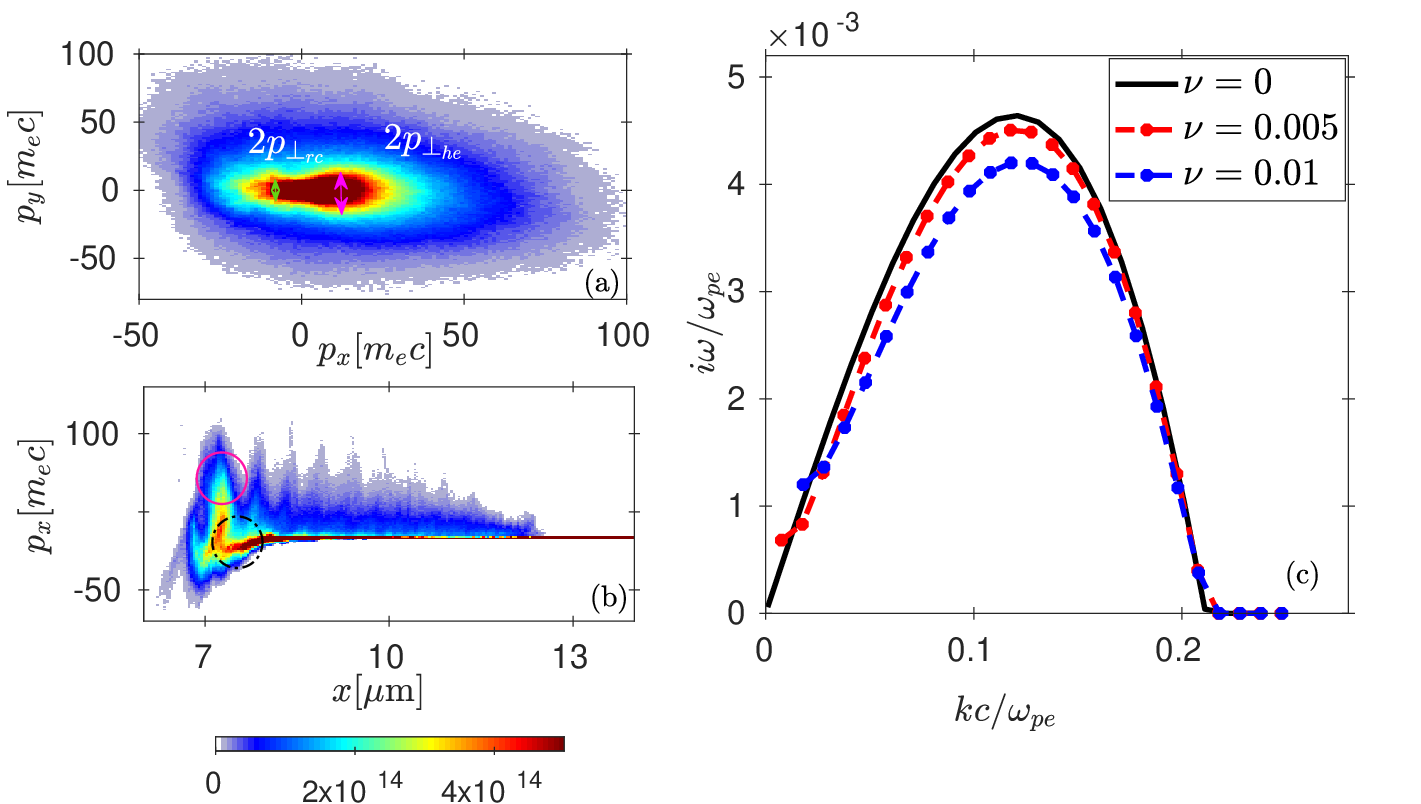}
\caption{Panel (a) shows the electron momenta { $(p_x$-$p_y)$} phase space during the linear phase of Weibel instability development (at $t=45 $ fs). The thermal spread of the counter-streaming electron streams is shown by the colored arrows respectively. \textcolor{black}{Panel (b) shows the electron  phase space $(x$-$p_x)$ at the same time. The solid circle marks the hot-electrons while the dot-dashed circle marks the return current component.} Panel (c) shows the growth rate of Weibel instability with and without collisions from the linear instability analysis. \textcolor{black}{The parameters are extracted from fitting the MJ curve on (a) for,  $n_{he}/n_{rc}=0.3$.} 
The collision frequency is varied from $\nu=\nu_0/\omega_{pe}=[0,0.005,0.01]$.}
\label{fig_MJ}
\end{figure}

\begin{table}[b]
\caption{\label{Appendix_table}%
The growth rate of Weibel instability with different density ratio. The parameters $\mu$'s and $\beta$'s are extracted from Fig.\ref{fig_MJ}(a) by fitting a Maxwell-J\"uttner type distribution function. Both \textcolor{black}{growth rates} collisionless ($\delta^0$) and collisional ($\delta^{0.01}$), corresponding to the collision frequency ($\nu=0.01$) are normalized by the plasma frequency.}%

\begin{tabular}{|c|c|c|c|c|}
\hline 
&$\mu_{he}(\mu_{rc})$ & $n_{he}/n_{rc}$ & $\beta_{he}(-\beta_{rc})$ & $\delta^0 (\delta^{0.01})$ \\ 
\hline 
Fitted MJ            &0.656(0.792)   & 0.1 & 0.485 (0.0461) & 0.0013 (0.0007) \\
\cline{2-5}
distribution         &0.656(0.792)   & 0.2 & 0.485 (0.0461) & 0.0030 (0.0024) \\ 
\cline{2-5}
from                 & 0.656(0.792)  & 0.3 & 0.485 (0.0461) & 0.0046 (0.0042) \\ 
\cline{2-5}
Fig.\ref{fig_MJ}(a)  & 0.656(0.792)  & 0.4 & 0.485 (0.0461) & 0.0060 (0.0057) \\ 
\cline{2-5} 
                     & 0.656(0.792)  & 0.5 & 0.485 (0.0461) & 0.0072 (0.0070) \\ 
\hline
Bret \emph{et al}.\cite{Bret_2010_exact}& 0.37(0.125)  & 0.3 & 0.999 (0.15) & 0.0035 (0.0031) \\ 
   \hline 
\end{tabular} 
\end{table} 

The comparison of the theoretical growth rates with simulation results in our case is not straightforward since parameters viz. temperature, momentum and the densities of both streams are not exactly defined. For this we extract the parallel and perpendicular momenta of the two streams are from PIC simulations. Fig.\ref{fig_MJ}(a) shows the electron momentum space during the linear phase (at $t=45$ fs) of the WI/CFI development~\cite{Tzoufras:2006}. A relativistic Maxwell-J{\"u}ttner distribution function as a function of $p_{\parallel}$ (in $x$-direction) and $ p_{\perp}$(in $y$-direction), for two drifting streams is then fitted over the data of electron's momentum distribution from the simulation shown in Fig.\ref{fig_MJ}(a) and the coefficients ($\mu_i,\beta_i$) are predicted with $95\%$ confidence bounds. Using this fitted distribution function we calculate the growth rates and these are summarised in Table~\ref{Appendix_table} for collision frequency $\nu=0.01$. The hot electron and return current density ratio is varied  and for a suitable density ratio ($n_{he}/n_{rc}$), one can find a good agreement between PIC simulation results and theoretical growth rates as shows in Table~\ref{table_growth_rate}. \textcolor{black}{Fig.\ref{fig_MJ}(b) shows the electron $(x$-$p_x)$ distribution function at $t=45$ fs. The solid circle marks the hot-electrons component of electrons while the dot-dashed black circle marks the return current component from which one can estimate the density ratio of the two streaming components to be in the range $[0.1-0.3]$.} The last row of Table~\ref{Appendix_table} shows the growth rate estimation by using the criteria for transverse and longitudinal momentum spread from Ref.\cite{Bret_2010_exact}. \textcolor{black}{On taking $p_{\parallel}^{he}\sim 59.7m_ec, n_{he}/n_{rc}=0.3$ while the $p_{\perp}^{he}\sim 18\, m_e c,\, p_{\perp}^{rc}\sim 4 m_e c$.} The theoretical growth rates are smaller than the PIC simulation values \textcolor{black}{in Table.~\ref{table_growth_rate}, pointing towards an enhanced Weibel instability growth rate due to trapped electrons in the downstream of electrostatic shock~\cite{Stockem:2014aa}}. Nevertheless they capture the trend of reduction in the growth rate due to collisions.


\begin{thebibliography}{44}%
\makeatletter
\providecommand \@ifxundefined [1]{%
 \@ifx{#1\undefined}
}%
\providecommand \@ifnum [1]{%
 \ifnum #1\expandafter \@firstoftwo
 \else \expandafter \@secondoftwo
 \fi
}%
\providecommand \@ifx [1]{%
 \ifx #1\expandafter \@firstoftwo
 \else \expandafter \@secondoftwo
 \fi
}%
\providecommand \natexlab [1]{#1}%
\providecommand \enquote  [1]{``#1''}%
\providecommand \bibnamefont  [1]{#1}%
\providecommand \bibfnamefont [1]{#1}%
\providecommand \citenamefont [1]{#1}%
\providecommand \href@noop [0]{\@secondoftwo}%
\providecommand \href [0]{\begingroup \@sanitize@url \@href}%
\providecommand \@href[1]{\@@startlink{#1}\@@href}%
\providecommand \@@href[1]{\endgroup#1\@@endlink}%
\providecommand \@sanitize@url [0]{\catcode `\\12\catcode `\$12\catcode
  `\&12\catcode `\#12\catcode `\^12\catcode `\_12\catcode `\%12\relax}%
\providecommand \@@startlink[1]{}%
\providecommand \@@endlink[0]{}%
\providecommand \url  [0]{\begingroup\@sanitize@url \@url }%
\providecommand \@url [1]{\endgroup\@href {#1}{\urlprefix }}%
\providecommand \urlprefix  [0]{URL }%
\providecommand \Eprint [0]{\href }%
\providecommand \doibase [0]{http://dx.doi.org/}%
\providecommand \selectlanguage [0]{\@gobble}%
\providecommand \bibinfo  [0]{\@secondoftwo}%
\providecommand \bibfield  [0]{\@secondoftwo}%
\providecommand \translation [1]{[#1]}%
\providecommand \BibitemOpen [0]{}%
\providecommand \bibitemStop [0]{}%
\providecommand \bibitemNoStop [0]{.\EOS\space}%
\providecommand \EOS [0]{\spacefactor3000\relax}%
\providecommand \BibitemShut  [1]{\csname bibitem#1\endcsname}%
\let\auto@bib@innerbib\@empty
\bibitem [{\citenamefont {Blandford}\ and\ \citenamefont
  {Eichler}(1987)}]{Blandford:1987aa}%
  \BibitemOpen
  \bibfield  {author} {\bibinfo {author} {\bibfnamefont {R.}~\bibnamefont
  {Blandford}}\ and\ \bibinfo {author} {\bibfnamefont {D.}~\bibnamefont
  {Eichler}},\ }\bibfield  {title} {\enquote {\bibinfo {title} {Particle
  acceleration at astrophysical shocks: A theory of cosmic ray origin},}\
  }\href {\doibase http://dx.doi.org/10.1016/0370-1573(87)90134-7} {\bibfield
  {journal} {\bibinfo  {journal} {Physics Reports}\ }\textbf {\bibinfo {volume}
  {154}},\ \bibinfo {pages} {1 -- 75} (\bibinfo {year} {1987})}\BibitemShut
  {NoStop}%
\bibitem [{\citenamefont {Spitkovsky}(2008)}]{Spitkovsky:2008aa}%
  \BibitemOpen
  \bibfield  {author} {\bibinfo {author} {\bibfnamefont {A.}~\bibnamefont
  {Spitkovsky}},\ }\bibfield  {title} {\enquote {\bibinfo {title} {Particle
  acceleration in relativistic collisionless shocks: Fermi process at last?}}\
  }\href {\doibase 10.1086/590248} {\bibfield  {journal} {\bibinfo  {journal}
  {The Astrophysical Journal}\ }\textbf {\bibinfo {volume} {682}},\ \bibinfo
  {pages} {L5--L8} (\bibinfo {year} {2008})}\BibitemShut {NoStop}%
\bibitem [{\citenamefont {Spitkovsky}(2007)}]{Spitkovsky:2007aa}%
  \BibitemOpen
  \bibfield  {author} {\bibinfo {author} {\bibfnamefont {A.}~\bibnamefont
  {Spitkovsky}},\ }\bibfield  {title} {\enquote {\bibinfo {title} {On the
  structure of relativistic collisionless shocks in electron-ion plasmas},}\
  }\href {\doibase 10.1086/527374} {\bibfield  {journal} {\bibinfo  {journal}
  {The Astrophysical Journal}\ }\textbf {\bibinfo {volume} {673}},\ \bibinfo
  {pages} {L39--L42} (\bibinfo {year} {2007})}\BibitemShut {NoStop}%
\bibitem [{\citenamefont {Romagnani}\ \emph {et~al.}(2008)\citenamefont
  {Romagnani}, \citenamefont {Bulanov}, \citenamefont {Borghesi}, \citenamefont
  {Audebert}, \citenamefont {Gauthier}, \citenamefont {L\"owenbr\"uck},
  \citenamefont {Mackinnon}, \citenamefont {Patel}, \citenamefont {Pretzler},
  \citenamefont {Toncian},\ and\ \citenamefont {Willi}}]{Romagnani:2008aa}%
  \BibitemOpen
  \bibfield  {author} {\bibinfo {author} {\bibfnamefont {L.}~\bibnamefont
  {Romagnani}}, \bibinfo {author} {\bibfnamefont {S.~V.}\ \bibnamefont
  {Bulanov}}, \bibinfo {author} {\bibfnamefont {M.}~\bibnamefont {Borghesi}},
  \bibinfo {author} {\bibfnamefont {P.}~\bibnamefont {Audebert}}, \bibinfo
  {author} {\bibfnamefont {J.~C.}\ \bibnamefont {Gauthier}}, \bibinfo {author}
  {\bibfnamefont {K.}~\bibnamefont {L\"owenbr\"uck}}, \bibinfo {author}
  {\bibfnamefont {A.~J.}\ \bibnamefont {Mackinnon}}, \bibinfo {author}
  {\bibfnamefont {P.}~\bibnamefont {Patel}}, \bibinfo {author} {\bibfnamefont
  {G.}~\bibnamefont {Pretzler}}, \bibinfo {author} {\bibfnamefont
  {T.}~\bibnamefont {Toncian}}, \ and\ \bibinfo {author} {\bibfnamefont
  {O.}~\bibnamefont {Willi}},\ }\bibfield  {title} {\enquote {\bibinfo {title}
  {Observation of collisionless shocks in laser-plasma experiments},}\ }\href
  {\doibase 10.1103/PhysRevLett.101.025004} {\bibfield  {journal} {\bibinfo
  {journal} {Phys. Rev. Lett.}\ }\textbf {\bibinfo {volume} {101}},\ \bibinfo
  {pages} {025004} (\bibinfo {year} {2008})}\BibitemShut {NoStop}%
\bibitem [{\citenamefont {Silva}\ \emph {et~al.}(2004)\citenamefont {Silva},
  \citenamefont {Marti}, \citenamefont {Davies}, \citenamefont {Fonseca},
  \citenamefont {Ren}, \citenamefont {Tsung},\ and\ \citenamefont
  {Mori}}]{Silva:2004aa}%
  \BibitemOpen
  \bibfield  {author} {\bibinfo {author} {\bibfnamefont {L.~O.}\ \bibnamefont
  {Silva}}, \bibinfo {author} {\bibfnamefont {M.}~\bibnamefont {Marti}},
  \bibinfo {author} {\bibfnamefont {J.~R.}\ \bibnamefont {Davies}}, \bibinfo
  {author} {\bibfnamefont {R.~A.}\ \bibnamefont {Fonseca}}, \bibinfo {author}
  {\bibfnamefont {C.}~\bibnamefont {Ren}}, \bibinfo {author} {\bibfnamefont
  {F.~S.}\ \bibnamefont {Tsung}}, \ and\ \bibinfo {author} {\bibfnamefont
  {W.~B.}\ \bibnamefont {Mori}},\ }\bibfield  {title} {\enquote {\bibinfo
  {title} {Proton shock acceleration in laser-plasma interactions},}\ }\href
  {\doibase 10.1103/PhysRevLett.92.015002} {\bibfield  {journal} {\bibinfo
  {journal} {Phys. Rev. Lett.}\ }\textbf {\bibinfo {volume} {92}},\ \bibinfo
  {pages} {015002} (\bibinfo {year} {2004})}\BibitemShut {NoStop}%
\bibitem [{\citenamefont {Haberberger}\ \emph {et~al.}(2012)\citenamefont
  {Haberberger}, \citenamefont {Tochitsky}, \citenamefont {Fiuza},
  \citenamefont {Gong}, \citenamefont {Fonseca}, \citenamefont {Silva},
  \citenamefont {Mori},\ and\ \citenamefont {Joshi}}]{Haberberger:2012aa}%
  \BibitemOpen
  \bibfield  {author} {\bibinfo {author} {\bibfnamefont {D.}~\bibnamefont
  {Haberberger}}, \bibinfo {author} {\bibfnamefont {S.}~\bibnamefont
  {Tochitsky}}, \bibinfo {author} {\bibfnamefont {F.}~\bibnamefont {Fiuza}},
  \bibinfo {author} {\bibfnamefont {C.}~\bibnamefont {Gong}}, \bibinfo {author}
  {\bibfnamefont {R.~A.}\ \bibnamefont {Fonseca}}, \bibinfo {author}
  {\bibfnamefont {L.~O.}\ \bibnamefont {Silva}}, \bibinfo {author}
  {\bibfnamefont {W.~B.}\ \bibnamefont {Mori}}, \ and\ \bibinfo {author}
  {\bibfnamefont {C.}~\bibnamefont {Joshi}},\ }\bibfield  {title} {\enquote
  {\bibinfo {title} {Collisionless shocks in laser-produced plasma generate
  monoenergetic high-energy proton beams},}\ }\href
  {http://dx.doi.org/10.1038/nphys2130} {\bibfield  {journal} {\bibinfo
  {journal} {Nat Phys}\ }\textbf {\bibinfo {volume} {8}},\ \bibinfo {pages}
  {95--99} (\bibinfo {year} {2012})}\BibitemShut {NoStop}%
\bibitem [{\citenamefont {Fiuza}\ \emph
  {et~al.}(2012{\natexlab{a}})\citenamefont {Fiuza}, \citenamefont {Stockem},
  \citenamefont {Boella}, \citenamefont {Fonseca}, \citenamefont {Silva},
  \citenamefont {Haberberger}, \citenamefont {Tochitsky}, \citenamefont {Gong},
  \citenamefont {Mori},\ and\ \citenamefont {Joshi}}]{Fiuza:2012ab}%
  \BibitemOpen
  \bibfield  {author} {\bibinfo {author} {\bibfnamefont {F.}~\bibnamefont
  {Fiuza}}, \bibinfo {author} {\bibfnamefont {A.}~\bibnamefont {Stockem}},
  \bibinfo {author} {\bibfnamefont {E.}~\bibnamefont {Boella}}, \bibinfo
  {author} {\bibfnamefont {R.~A.}\ \bibnamefont {Fonseca}}, \bibinfo {author}
  {\bibfnamefont {L.~O.}\ \bibnamefont {Silva}}, \bibinfo {author}
  {\bibfnamefont {D.}~\bibnamefont {Haberberger}}, \bibinfo {author}
  {\bibfnamefont {S.}~\bibnamefont {Tochitsky}}, \bibinfo {author}
  {\bibfnamefont {C.}~\bibnamefont {Gong}}, \bibinfo {author} {\bibfnamefont
  {W.~B.}\ \bibnamefont {Mori}}, \ and\ \bibinfo {author} {\bibfnamefont
  {C.}~\bibnamefont {Joshi}},\ }\bibfield  {title} {\enquote {\bibinfo {title}
  {Laser-driven shock acceleration of monoenergetic ion beams},}\ }\href
  {\doibase 10.1103/physrevlett.109.215001} {\bibfield  {journal} {\bibinfo
  {journal} {Physical Review Letters}\ }\textbf {\bibinfo {volume} {109}}
  (\bibinfo {year} {2012}{\natexlab{a}}),\
  10.1103/physrevlett.109.215001}\BibitemShut {NoStop}%
\bibitem [{\citenamefont {Fiuza}\ \emph {et~al.}(2013)\citenamefont {Fiuza},
  \citenamefont {Stockem}, \citenamefont {Boella}, \citenamefont {Fonseca},
  \citenamefont {Silva}, \citenamefont {Haberberger}, \citenamefont
  {Tochitsky}, \citenamefont {Mori},\ and\ \citenamefont
  {Joshi}}]{Fiuza:2013aa}%
  \BibitemOpen
  \bibfield  {author} {\bibinfo {author} {\bibfnamefont {F.}~\bibnamefont
  {Fiuza}}, \bibinfo {author} {\bibfnamefont {A.}~\bibnamefont {Stockem}},
  \bibinfo {author} {\bibfnamefont {E.}~\bibnamefont {Boella}}, \bibinfo
  {author} {\bibfnamefont {R.~A.}\ \bibnamefont {Fonseca}}, \bibinfo {author}
  {\bibfnamefont {L.~O.}\ \bibnamefont {Silva}}, \bibinfo {author}
  {\bibfnamefont {D.}~\bibnamefont {Haberberger}}, \bibinfo {author}
  {\bibfnamefont {S.}~\bibnamefont {Tochitsky}}, \bibinfo {author}
  {\bibfnamefont {W.~B.}\ \bibnamefont {Mori}}, \ and\ \bibinfo {author}
  {\bibfnamefont {C.}~\bibnamefont {Joshi}},\ }\bibfield  {title} {\enquote
  {\bibinfo {title} {Ion acceleration from laser-driven electrostatic
  shocks},}\ }\href {\doibase 10.1063/1.4801526} {\bibfield  {journal}
  {\bibinfo  {journal} {Physics of Plasmas}\ }\textbf {\bibinfo {volume}
  {20}},\ \bibinfo {pages} {056304} (\bibinfo {year} {2013})}\BibitemShut
  {NoStop}%
\bibitem [{\citenamefont {Salamin}, \citenamefont {Harman},\ and\ \citenamefont
  {Keitel}(2008)}]{Salamin:2008aa}%
  \BibitemOpen
  \bibfield  {author} {\bibinfo {author} {\bibfnamefont {Y.~I.}\ \bibnamefont
  {Salamin}}, \bibinfo {author} {\bibfnamefont {Z.}~\bibnamefont {Harman}}, \
  and\ \bibinfo {author} {\bibfnamefont {C.~H.}\ \bibnamefont {Keitel}},\
  }\bibfield  {title} {\enquote {\bibinfo {title} {Direct high-power laser
  acceleration of ions for medical applications},}\ }\href {\doibase
  10.1103/PhysRevLett.100.155004} {\bibfield  {journal} {\bibinfo  {journal}
  {Phys. Rev. Lett.}\ }\textbf {\bibinfo {volume} {100}},\ \bibinfo {pages}
  {155004} (\bibinfo {year} {2008})}\BibitemShut {NoStop}%
\bibitem [{\citenamefont {Linz}\ and\ \citenamefont
  {Alonso}(2007)}]{Linz:2007aa}%
  \BibitemOpen
  \bibfield  {author} {\bibinfo {author} {\bibfnamefont {U.}~\bibnamefont
  {Linz}}\ and\ \bibinfo {author} {\bibfnamefont {J.}~\bibnamefont {Alonso}},\
  }\bibfield  {title} {\enquote {\bibinfo {title} {What will it take for laser
  driven proton accelerators to be applied to tumor therapy?}}\ }\href
  {\doibase 10.1103/PhysRevSTAB.10.094801} {\bibfield  {journal} {\bibinfo
  {journal} {Phys. Rev. ST Accel. Beams}\ }\textbf {\bibinfo {volume} {10}},\
  \bibinfo {pages} {094801} (\bibinfo {year} {2007})}\BibitemShut {NoStop}%
\bibitem [{\citenamefont {Turrell}, \citenamefont {Sherlock},\ and\
  \citenamefont {Rose}(2015)}]{Turrell:2015aa}%
  \BibitemOpen
  \bibfield  {author} {\bibinfo {author} {\bibfnamefont {A.~E.}\ \bibnamefont
  {Turrell}}, \bibinfo {author} {\bibfnamefont {M.}~\bibnamefont {Sherlock}}, \
  and\ \bibinfo {author} {\bibfnamefont {S.~J.}\ \bibnamefont {Rose}},\
  }\bibfield  {title} {\enquote {\bibinfo {title} {Ultrafast collisional ion
  heating by electrostatic shocks},}\ }\href {\doibase 10.1038/ncomms9905}
  {\bibfield  {journal} {\bibinfo  {journal} {Nature Communications}\ }\textbf
  {\bibinfo {volume} {6}},\ \bibinfo {pages} {8905} (\bibinfo {year}
  {2015})}\BibitemShut {NoStop}%
\bibitem [{\citenamefont {Forslund}\ and\ \citenamefont
  {Shonk}(1970)}]{Forslund:1970aa}%
  \BibitemOpen
  \bibfield  {author} {\bibinfo {author} {\bibfnamefont {D.~W.}\ \bibnamefont
  {Forslund}}\ and\ \bibinfo {author} {\bibfnamefont {C.~R.}\ \bibnamefont
  {Shonk}},\ }\bibfield  {title} {\enquote {\bibinfo {title} {Formation and
  structure of electrostatic collisionless shocks},}\ }\href {\doibase
  10.1103/PhysRevLett.25.1699} {\bibfield  {journal} {\bibinfo  {journal}
  {Phys. Rev. Lett.}\ }\textbf {\bibinfo {volume} {25}},\ \bibinfo {pages}
  {1699--1702} (\bibinfo {year} {1970})}\BibitemShut {NoStop}%
\bibitem [{\citenamefont {Kato}\ and\ \citenamefont
  {Takabe}(2010)}]{Kato:2010aa}%
  \BibitemOpen
  \bibfield  {author} {\bibinfo {author} {\bibfnamefont {T.~N.}\ \bibnamefont
  {Kato}}\ and\ \bibinfo {author} {\bibfnamefont {H.}~\bibnamefont {Takabe}},\
  }\bibfield  {title} {\enquote {\bibinfo {title} {Electrostatic and
  electromagnetic instabilities associated with electrostatic shocks:
  Two-dimensional particle-in-cell simulation},}\ }\href {\doibase
  10.1063/1.3372138} {\bibfield  {journal} {\bibinfo  {journal} {Physics of
  Plasmas}\ }\textbf {\bibinfo {volume} {17}},\ \bibinfo {pages} {032114}
  (\bibinfo {year} {2010})}\BibitemShut {NoStop}%
\bibitem [{\citenamefont {Stockem}\ \emph
  {et~al.}(2014{\natexlab{a}})\citenamefont {Stockem}, \citenamefont {Fiuza},
  \citenamefont {Bret}, \citenamefont {Fonseca},\ and\ \citenamefont
  {Silva}}]{Stockem:2014ab}%
  \BibitemOpen
  \bibfield  {author} {\bibinfo {author} {\bibfnamefont {A.}~\bibnamefont
  {Stockem}}, \bibinfo {author} {\bibfnamefont {F.}~\bibnamefont {Fiuza}},
  \bibinfo {author} {\bibfnamefont {A.}~\bibnamefont {Bret}}, \bibinfo {author}
  {\bibfnamefont {R.~A.}\ \bibnamefont {Fonseca}}, \ and\ \bibinfo {author}
  {\bibfnamefont {L.~O.}\ \bibnamefont {Silva}},\ }\bibfield  {title} {\enquote
  {\bibinfo {title} {Exploring the nature of collisionless shocks under
  laboratory conditions},}\ }\href {http://dx.doi.org/10.1038/srep03934}
  {\bibfield  {journal} {\bibinfo  {journal} {Scientific Reports}\ }\textbf
  {\bibinfo {volume} {4}},\ \bibinfo {pages} {3934} (\bibinfo {year}
  {2014}{\natexlab{a}})}\BibitemShut {NoStop}%
\bibitem [{\citenamefont {Bret}\ \emph {et~al.}(2013)\citenamefont {Bret},
  \citenamefont {Stockem}, \citenamefont {Fiuza}, \citenamefont {Ruyer},
  \citenamefont {Gremillet}, \citenamefont {Narayan},\ and\ \citenamefont
  {Silva}}]{Bret:2013aa}%
  \BibitemOpen
  \bibfield  {author} {\bibinfo {author} {\bibfnamefont {A.}~\bibnamefont
  {Bret}}, \bibinfo {author} {\bibfnamefont {A.}~\bibnamefont {Stockem}},
  \bibinfo {author} {\bibfnamefont {F.}~\bibnamefont {Fiuza}}, \bibinfo
  {author} {\bibfnamefont {C.}~\bibnamefont {Ruyer}}, \bibinfo {author}
  {\bibfnamefont {L.}~\bibnamefont {Gremillet}}, \bibinfo {author}
  {\bibfnamefont {R.}~\bibnamefont {Narayan}}, \ and\ \bibinfo {author}
  {\bibfnamefont {L.~O.}\ \bibnamefont {Silva}},\ }\bibfield  {title} {\enquote
  {\bibinfo {title} {Collisionless shock formation, spontaneous electromagnetic
  fluctuations, and streaming instabilities},}\ }\href {\doibase
  10.1063/1.4798541} {\bibfield  {journal} {\bibinfo  {journal} {Physics of
  Plasmas}\ }\textbf {\bibinfo {volume} {20}},\ \bibinfo {pages} {042102}
  (\bibinfo {year} {2013})}\BibitemShut {NoStop}%
\bibitem [{\citenamefont {Ryutov}\ \emph {et~al.}(2014)\citenamefont {Ryutov},
  \citenamefont {Fiuza}, \citenamefont {Huntington}, \citenamefont {Ross},\
  and\ \citenamefont {Park}}]{Ryutov:2014aa}%
  \BibitemOpen
  \bibfield  {author} {\bibinfo {author} {\bibfnamefont {D.~D.}\ \bibnamefont
  {Ryutov}}, \bibinfo {author} {\bibfnamefont {F.}~\bibnamefont {Fiuza}},
  \bibinfo {author} {\bibfnamefont {C.~M.}\ \bibnamefont {Huntington}},
  \bibinfo {author} {\bibfnamefont {J.~S.}\ \bibnamefont {Ross}}, \ and\
  \bibinfo {author} {\bibfnamefont {H.~S.}\ \bibnamefont {Park}},\ }\bibfield
  {title} {\enquote {\bibinfo {title} {Collisional effects in the ion weibel
  instability for two counter-propagating plasma streams},}\ }\href {\doibase
  10.1063/1.4867062} {\bibfield  {journal} {\bibinfo  {journal} {Physics of
  Plasmas}\ }\textbf {\bibinfo {volume} {21}},\ \bibinfo {pages} {032701}
  (\bibinfo {year} {2014})}\BibitemShut {NoStop}%
\bibitem [{\citenamefont {Ross}\ \emph {et~al.}(2017)\citenamefont {Ross},
  \citenamefont {Higginson}, \citenamefont {Ryutov}, \citenamefont {Fiuza},
  \citenamefont {Hatarik}, \citenamefont {Huntington}, \citenamefont
  {Kalantar}, \citenamefont {Link}, \citenamefont {Pollock}, \citenamefont
  {Remington}, \citenamefont {Rinderknecht}, \citenamefont {Swadling},
  \citenamefont {Turnbull}, \citenamefont {Weber}, \citenamefont {Wilks},
  \citenamefont {Froula}, \citenamefont {Rosenberg}, \citenamefont {Morita},
  \citenamefont {Sakawa}, \citenamefont {Takabe}, \citenamefont {Drake},
  \citenamefont {Kuranz}, \citenamefont {Gregori}, \citenamefont {Meinecke},
  \citenamefont {Levy}, \citenamefont {Koenig}, \citenamefont {Spitkovsky},
  \citenamefont {Petrasso}, \citenamefont {Li}, \citenamefont {Sio},
  \citenamefont {Lahmann}, \citenamefont {Zylstra},\ and\ \citenamefont
  {Park}}]{Ross:2017aa}%
  \BibitemOpen
  \bibfield  {author} {\bibinfo {author} {\bibfnamefont {J.~S.}\ \bibnamefont
  {Ross}}, \bibinfo {author} {\bibfnamefont {D.~P.}\ \bibnamefont {Higginson}},
  \bibinfo {author} {\bibfnamefont {D.}~\bibnamefont {Ryutov}}, \bibinfo
  {author} {\bibfnamefont {F.}~\bibnamefont {Fiuza}}, \bibinfo {author}
  {\bibfnamefont {R.}~\bibnamefont {Hatarik}}, \bibinfo {author} {\bibfnamefont
  {C.~M.}\ \bibnamefont {Huntington}}, \bibinfo {author} {\bibfnamefont
  {D.~H.}\ \bibnamefont {Kalantar}}, \bibinfo {author} {\bibfnamefont
  {A.}~\bibnamefont {Link}}, \bibinfo {author} {\bibfnamefont {B.~B.}\
  \bibnamefont {Pollock}}, \bibinfo {author} {\bibfnamefont {B.~A.}\
  \bibnamefont {Remington}}, \bibinfo {author} {\bibfnamefont {H.~G.}\
  \bibnamefont {Rinderknecht}}, \bibinfo {author} {\bibfnamefont {G.~F.}\
  \bibnamefont {Swadling}}, \bibinfo {author} {\bibfnamefont {D.~P.}\
  \bibnamefont {Turnbull}}, \bibinfo {author} {\bibfnamefont {S.}~\bibnamefont
  {Weber}}, \bibinfo {author} {\bibfnamefont {S.}~\bibnamefont {Wilks}},
  \bibinfo {author} {\bibfnamefont {D.~H.}\ \bibnamefont {Froula}}, \bibinfo
  {author} {\bibfnamefont {M.~J.}\ \bibnamefont {Rosenberg}}, \bibinfo {author}
  {\bibfnamefont {T.}~\bibnamefont {Morita}}, \bibinfo {author} {\bibfnamefont
  {Y.}~\bibnamefont {Sakawa}}, \bibinfo {author} {\bibfnamefont
  {H.}~\bibnamefont {Takabe}}, \bibinfo {author} {\bibfnamefont {R.~P.}\
  \bibnamefont {Drake}}, \bibinfo {author} {\bibfnamefont {C.}~\bibnamefont
  {Kuranz}}, \bibinfo {author} {\bibfnamefont {G.}~\bibnamefont {Gregori}},
  \bibinfo {author} {\bibfnamefont {J.}~\bibnamefont {Meinecke}}, \bibinfo
  {author} {\bibfnamefont {M.~C.}\ \bibnamefont {Levy}}, \bibinfo {author}
  {\bibfnamefont {M.}~\bibnamefont {Koenig}}, \bibinfo {author} {\bibfnamefont
  {A.}~\bibnamefont {Spitkovsky}}, \bibinfo {author} {\bibfnamefont {R.~D.}\
  \bibnamefont {Petrasso}}, \bibinfo {author} {\bibfnamefont {C.~K.}\
  \bibnamefont {Li}}, \bibinfo {author} {\bibfnamefont {H.}~\bibnamefont
  {Sio}}, \bibinfo {author} {\bibfnamefont {B.}~\bibnamefont {Lahmann}},
  \bibinfo {author} {\bibfnamefont {A.~B.}\ \bibnamefont {Zylstra}}, \ and\
  \bibinfo {author} {\bibfnamefont {H.-S.}\ \bibnamefont {Park}},\ }\bibfield
  {title} {\enquote {\bibinfo {title} {Transition from collisional to
  collisionless regimes in interpenetrating plasma flows on the national
  ignition facility},}\ }\href {\doibase 10.1103/PhysRevLett.118.185003}
  {\bibfield  {journal} {\bibinfo  {journal} {Phys. Rev. Lett.}\ }\textbf
  {\bibinfo {volume} {118}},\ \bibinfo {pages} {185003} (\bibinfo {year}
  {2017})}\BibitemShut {NoStop}%
\bibitem [{\citenamefont {Huntington}\ \emph {et~al.}(2015)\citenamefont
  {Huntington}, \citenamefont {Fiuza}, \citenamefont {Ross}, \citenamefont
  {Zylstra}, \citenamefont {Drake}, \citenamefont {Froula}, \citenamefont
  {Gregori}, \citenamefont {Kugland}, \citenamefont {Kuranz}, \citenamefont
  {Levy}, \citenamefont {Li}, \citenamefont {Meinecke}, \citenamefont {Morita},
  \citenamefont {Petrasso}, \citenamefont {Plechaty}, \citenamefont
  {Remington}, \citenamefont {Ryutov}, \citenamefont {Sakawa}, \citenamefont
  {Spitkovsky}, \citenamefont {Takabe},\ and\ \citenamefont
  {Park}}]{Huntington:2015aa}%
  \BibitemOpen
  \bibfield  {author} {\bibinfo {author} {\bibfnamefont {C.~M.}\ \bibnamefont
  {Huntington}}, \bibinfo {author} {\bibfnamefont {F.}~\bibnamefont {Fiuza}},
  \bibinfo {author} {\bibfnamefont {J.~S.}\ \bibnamefont {Ross}}, \bibinfo
  {author} {\bibfnamefont {A.~B.}\ \bibnamefont {Zylstra}}, \bibinfo {author}
  {\bibfnamefont {R.~P.}\ \bibnamefont {Drake}}, \bibinfo {author}
  {\bibfnamefont {D.~H.}\ \bibnamefont {Froula}}, \bibinfo {author}
  {\bibfnamefont {G.}~\bibnamefont {Gregori}}, \bibinfo {author} {\bibfnamefont
  {N.~L.}\ \bibnamefont {Kugland}}, \bibinfo {author} {\bibfnamefont {C.~C.}\
  \bibnamefont {Kuranz}}, \bibinfo {author} {\bibfnamefont {M.~C.}\
  \bibnamefont {Levy}}, \bibinfo {author} {\bibfnamefont {C.~K.}\ \bibnamefont
  {Li}}, \bibinfo {author} {\bibfnamefont {J.}~\bibnamefont {Meinecke}},
  \bibinfo {author} {\bibfnamefont {T.}~\bibnamefont {Morita}}, \bibinfo
  {author} {\bibfnamefont {R.}~\bibnamefont {Petrasso}}, \bibinfo {author}
  {\bibfnamefont {C.}~\bibnamefont {Plechaty}}, \bibinfo {author}
  {\bibfnamefont {B.~A.}\ \bibnamefont {Remington}}, \bibinfo {author}
  {\bibfnamefont {D.~D.}\ \bibnamefont {Ryutov}}, \bibinfo {author}
  {\bibfnamefont {Y.}~\bibnamefont {Sakawa}}, \bibinfo {author} {\bibfnamefont
  {A.}~\bibnamefont {Spitkovsky}}, \bibinfo {author} {\bibfnamefont
  {H.}~\bibnamefont {Takabe}}, \ and\ \bibinfo {author} {\bibfnamefont {H.-S.}\
  \bibnamefont {Park}},\ }\bibfield  {title} {\enquote {\bibinfo {title}
  {Observation of magnetic field generation via the weibel instability in
  interpenetrating plasma flows},}\ }\href {\doibase 10.1038/nphys3178}
  {\bibfield  {journal} {\bibinfo  {journal} {Nature Physics}\ }\textbf
  {\bibinfo {volume} {11}},\ \bibinfo {pages} {173--176} (\bibinfo {year}
  {2015})}\BibitemShut {NoStop}%
\bibitem [{\citenamefont {Park}\ \emph {et~al.}(2015)\citenamefont {Park},
  \citenamefont {Huntington}, \citenamefont {Fiuza}, \citenamefont {Drake},
  \citenamefont {Froula}, \citenamefont {Gregori}, \citenamefont {Koenig},
  \citenamefont {Kugland}, \citenamefont {Kuranz}, \citenamefont {Lamb},
  \citenamefont {Levy}, \citenamefont {Li}, \citenamefont {Meinecke},
  \citenamefont {Morita}, \citenamefont {Petrasso}, \citenamefont {Pollock},
  \citenamefont {Remington}, \citenamefont {Rinderknecht}, \citenamefont
  {Rosenberg}, \citenamefont {Ross}, \citenamefont {Ryutov}, \citenamefont
  {Sakawa}, \citenamefont {Spitkovsky}, \citenamefont {Takabe}, \citenamefont
  {Turnbull}, \citenamefont {Tzeferacos}, \citenamefont {Weber},\ and\
  \citenamefont {Zylstra}}]{Park:2015aa}%
  \BibitemOpen
  \bibfield  {author} {\bibinfo {author} {\bibfnamefont {H.~S.}\ \bibnamefont
  {Park}}, \bibinfo {author} {\bibfnamefont {C.~M.}\ \bibnamefont
  {Huntington}}, \bibinfo {author} {\bibfnamefont {F.}~\bibnamefont {Fiuza}},
  \bibinfo {author} {\bibfnamefont {R.~P.}\ \bibnamefont {Drake}}, \bibinfo
  {author} {\bibfnamefont {D.~H.}\ \bibnamefont {Froula}}, \bibinfo {author}
  {\bibfnamefont {G.}~\bibnamefont {Gregori}}, \bibinfo {author} {\bibfnamefont
  {M.}~\bibnamefont {Koenig}}, \bibinfo {author} {\bibfnamefont {N.~L.}\
  \bibnamefont {Kugland}}, \bibinfo {author} {\bibfnamefont {C.~C.}\
  \bibnamefont {Kuranz}}, \bibinfo {author} {\bibfnamefont {D.~Q.}\
  \bibnamefont {Lamb}}, \bibinfo {author} {\bibfnamefont {M.~C.}\ \bibnamefont
  {Levy}}, \bibinfo {author} {\bibfnamefont {C.~K.}\ \bibnamefont {Li}},
  \bibinfo {author} {\bibfnamefont {J.}~\bibnamefont {Meinecke}}, \bibinfo
  {author} {\bibfnamefont {T.}~\bibnamefont {Morita}}, \bibinfo {author}
  {\bibfnamefont {R.~D.}\ \bibnamefont {Petrasso}}, \bibinfo {author}
  {\bibfnamefont {B.~B.}\ \bibnamefont {Pollock}}, \bibinfo {author}
  {\bibfnamefont {B.~A.}\ \bibnamefont {Remington}}, \bibinfo {author}
  {\bibfnamefont {H.~G.}\ \bibnamefont {Rinderknecht}}, \bibinfo {author}
  {\bibfnamefont {M.}~\bibnamefont {Rosenberg}}, \bibinfo {author}
  {\bibfnamefont {J.~S.}\ \bibnamefont {Ross}}, \bibinfo {author}
  {\bibfnamefont {D.~D.}\ \bibnamefont {Ryutov}}, \bibinfo {author}
  {\bibfnamefont {Y.}~\bibnamefont {Sakawa}}, \bibinfo {author} {\bibfnamefont
  {A.}~\bibnamefont {Spitkovsky}}, \bibinfo {author} {\bibfnamefont
  {H.}~\bibnamefont {Takabe}}, \bibinfo {author} {\bibfnamefont {D.~P.}\
  \bibnamefont {Turnbull}}, \bibinfo {author} {\bibfnamefont {P.}~\bibnamefont
  {Tzeferacos}}, \bibinfo {author} {\bibfnamefont {S.~V.}\ \bibnamefont
  {Weber}}, \ and\ \bibinfo {author} {\bibfnamefont {A.~B.}\ \bibnamefont
  {Zylstra}},\ }\bibfield  {title} {\enquote {\bibinfo {title} {Collisionless
  shock experiments with lasers and observation of weibel instabilities},}\
  }\href {\doibase 10.1063/1.4920959} {\bibfield  {journal} {\bibinfo
  {journal} {Physics of Plasmas}\ }\textbf {\bibinfo {volume} {22}},\ \bibinfo
  {pages} {056311} (\bibinfo {year} {2015})}\BibitemShut {NoStop}%
\bibitem [{\citenamefont {Ruyer}, \citenamefont {Gremillet},\ and\
  \citenamefont {Bonnaud}(2015)}]{Ruyer:2015aa}%
  \BibitemOpen
  \bibfield  {author} {\bibinfo {author} {\bibfnamefont {C.}~\bibnamefont
  {Ruyer}}, \bibinfo {author} {\bibfnamefont {L.}~\bibnamefont {Gremillet}}, \
  and\ \bibinfo {author} {\bibfnamefont {G.}~\bibnamefont {Bonnaud}},\
  }\bibfield  {title} {\enquote {\bibinfo {title} {Weibel-mediated
  collisionless shocks in laser-irradiated dense plasmas: Prevailing role of
  the electrons in generating the field fluctuations},}\ }\href {\doibase
  10.1063/1.4928096} {\bibfield  {journal} {\bibinfo  {journal} {Physics of
  Plasmas}\ }\textbf {\bibinfo {volume} {22}},\ \bibinfo {pages} {082107}
  (\bibinfo {year} {2015})}\BibitemShut {NoStop}%
\bibitem [{\citenamefont {Fiuza}\ \emph
  {et~al.}(2012{\natexlab{b}})\citenamefont {Fiuza}, \citenamefont {Fonseca},
  \citenamefont {Tonge}, \citenamefont {Mori},\ and\ \citenamefont
  {Silva}}]{Fiuza:2012aa}%
  \BibitemOpen
  \bibfield  {author} {\bibinfo {author} {\bibfnamefont {F.}~\bibnamefont
  {Fiuza}}, \bibinfo {author} {\bibfnamefont {R.~A.}\ \bibnamefont {Fonseca}},
  \bibinfo {author} {\bibfnamefont {J.}~\bibnamefont {Tonge}}, \bibinfo
  {author} {\bibfnamefont {W.~B.}\ \bibnamefont {Mori}}, \ and\ \bibinfo
  {author} {\bibfnamefont {L.~O.}\ \bibnamefont {Silva}},\ }\bibfield  {title}
  {\enquote {\bibinfo {title} {Weibel-instability-mediated collisionless shocks
  in the laboratory with ultraintense lasers},}\ }\href {\doibase
  10.1103/PhysRevLett.108.235004} {\bibfield  {journal} {\bibinfo  {journal}
  {Phys. Rev. Lett.}\ }\textbf {\bibinfo {volume} {108}},\ \bibinfo {pages}
  {235004} (\bibinfo {year} {2012}{\natexlab{b}})}\BibitemShut {NoStop}%
\bibitem [{\citenamefont {Weibel}(1959)}]{Weibel:1959aa}%
  \BibitemOpen
  \bibfield  {author} {\bibinfo {author} {\bibfnamefont {E.~S.}\ \bibnamefont
  {Weibel}},\ }\bibfield  {title} {\enquote {\bibinfo {title} {Spontaneously
  growing transverse waves in a plasma due to an anisotropic velocity
  distribution},}\ }\href {\doibase 10.1103/PhysRevLett.2.83} {\bibfield
  {journal} {\bibinfo  {journal} {Phys. Rev. Lett.}\ }\textbf {\bibinfo
  {volume} {2}},\ \bibinfo {pages} {83--84} (\bibinfo {year}
  {1959})}\BibitemShut {NoStop}%
\bibitem [{\citenamefont {Karmakar}\ \emph {et~al.}(2009)\citenamefont
  {Karmakar}, \citenamefont {Kumar}, \citenamefont {Pukhov}, \citenamefont
  {Polomarov},\ and\ \citenamefont {Shvets}}]{Karmakar:2009aa}%
  \BibitemOpen
  \bibfield  {author} {\bibinfo {author} {\bibfnamefont {A.}~\bibnamefont
  {Karmakar}}, \bibinfo {author} {\bibfnamefont {N.}~\bibnamefont {Kumar}},
  \bibinfo {author} {\bibfnamefont {A.}~\bibnamefont {Pukhov}}, \bibinfo
  {author} {\bibfnamefont {O.}~\bibnamefont {Polomarov}}, \ and\ \bibinfo
  {author} {\bibfnamefont {G.}~\bibnamefont {Shvets}},\ }\bibfield  {title}
  {\enquote {\bibinfo {title} {Detailed particle-in-cell simulations on the
  transport of a relativistic electron beam in plasmas},}\ }\href {\doibase
  10.1103/PhysRevE.80.016401} {\bibfield  {journal} {\bibinfo  {journal} {Phys.
  Rev. E}\ }\textbf {\bibinfo {volume} {80}},\ \bibinfo {pages} {016401}
  (\bibinfo {year} {2009})}\BibitemShut {NoStop}%
\bibitem [{\citenamefont {Bret}, \citenamefont {Gremillet},\ and\ \citenamefont
  {Dieckmann}(2010)}]{Bret:2010aa}%
  \BibitemOpen
  \bibfield  {author} {\bibinfo {author} {\bibfnamefont {A.}~\bibnamefont
  {Bret}}, \bibinfo {author} {\bibfnamefont {L.}~\bibnamefont {Gremillet}}, \
  and\ \bibinfo {author} {\bibfnamefont {M.~E.}\ \bibnamefont {Dieckmann}},\
  }\bibfield  {title} {\enquote {\bibinfo {title} {Multidimensional electron
  beam-plasma instabilities in the relativistic regime},}\ }\href {\doibase
  10.1063/1.3514586} {\bibfield  {journal} {\bibinfo  {journal} {Physics of
  Plasmas}\ }\textbf {\bibinfo {volume} {17}},\ \bibinfo {pages} {120501}
  (\bibinfo {year} {2010})}\BibitemShut {NoStop}%
\bibitem [{\citenamefont {Bhadoria}\ and\ \citenamefont
  {Kumar}(2019)}]{Bhadoria2019}%
  \BibitemOpen
  \bibfield  {author} {\bibinfo {author} {\bibfnamefont {S.}~\bibnamefont
  {Bhadoria}}\ and\ \bibinfo {author} {\bibfnamefont {N.}~\bibnamefont
  {Kumar}},\ }\bibfield  {title} {\enquote {\bibinfo {title} {Collisionless
  shock acceleration of quasimonoenergetic ions in ultrarelativistic regime},}\
  }\href {\doibase 10.1103/PhysRevE.99.043205} {\bibfield  {journal} {\bibinfo
  {journal} {Phys. Rev. E}\ }\textbf {\bibinfo {volume} {99}},\ \bibinfo
  {pages} {043205} (\bibinfo {year} {2019})}\BibitemShut {NoStop}%
\bibitem [{\citenamefont {Arber}\ \emph {et~al.}(2015)\citenamefont {Arber},
  \citenamefont {Bennett}, \citenamefont {Brady}, \citenamefont
  {Lawrence-Douglas}, \citenamefont {Ramsay}, \citenamefont {Sircombe},
  \citenamefont {Gillies}, \citenamefont {Evans}, \citenamefont {Schmitz},
  \citenamefont {Bell},\ and\ \citenamefont {Ridgers}}]{Arber:2015aa}%
  \BibitemOpen
  \bibfield  {author} {\bibinfo {author} {\bibfnamefont {T.~D.}\ \bibnamefont
  {Arber}}, \bibinfo {author} {\bibfnamefont {K.}~\bibnamefont {Bennett}},
  \bibinfo {author} {\bibfnamefont {C.~S.}\ \bibnamefont {Brady}}, \bibinfo
  {author} {\bibfnamefont {A.}~\bibnamefont {Lawrence-Douglas}}, \bibinfo
  {author} {\bibfnamefont {M.~G.}\ \bibnamefont {Ramsay}}, \bibinfo {author}
  {\bibfnamefont {N.~J.}\ \bibnamefont {Sircombe}}, \bibinfo {author}
  {\bibfnamefont {P.}~\bibnamefont {Gillies}}, \bibinfo {author} {\bibfnamefont
  {R.~G.}\ \bibnamefont {Evans}}, \bibinfo {author} {\bibfnamefont
  {H.}~\bibnamefont {Schmitz}}, \bibinfo {author} {\bibfnamefont {A.~R.}\
  \bibnamefont {Bell}}, \ and\ \bibinfo {author} {\bibfnamefont {C.~P.}\
  \bibnamefont {Ridgers}},\ }\bibfield  {title} {\enquote {\bibinfo {title}
  {Contemporary particle-in-cell approach to laser-plasma modelling},}\ }\href
  {http://stacks.iop.org/0741-3335/57/i=11/a=113001} {\bibfield  {journal}
  {\bibinfo  {journal} {Plasma Physics and Controlled Fusion}\ }\textbf
  {\bibinfo {volume} {57}},\ \bibinfo {pages} {113001} (\bibinfo {year}
  {2015})}\BibitemShut {NoStop}%
\bibitem [{\citenamefont {Derouillat}\ \emph {et~al.}(2018)\citenamefont
  {Derouillat}, \citenamefont {Beck}, \citenamefont {P{\'e}rez}, \citenamefont
  {Vinci}, \citenamefont {Chiaramello}, \citenamefont {Grassi}, \citenamefont
  {Fl{\'e}}, \citenamefont {Bouchard}, \citenamefont {Plotnikov}, \citenamefont
  {Aunai}, \citenamefont {Dargent}, \citenamefont {Riconda},\ and\
  \citenamefont {Grech}}]{SMILEI-v4.1}%
  \BibitemOpen
  \bibfield  {author} {\bibinfo {author} {\bibfnamefont {J.}~\bibnamefont
  {Derouillat}}, \bibinfo {author} {\bibfnamefont {A.}~\bibnamefont {Beck}},
  \bibinfo {author} {\bibfnamefont {F.}~\bibnamefont {P{\'e}rez}}, \bibinfo
  {author} {\bibfnamefont {T.}~\bibnamefont {Vinci}}, \bibinfo {author}
  {\bibfnamefont {M.}~\bibnamefont {Chiaramello}}, \bibinfo {author}
  {\bibfnamefont {A.}~\bibnamefont {Grassi}}, \bibinfo {author} {\bibfnamefont
  {M.}~\bibnamefont {Fl{\'e}}}, \bibinfo {author} {\bibfnamefont
  {G.}~\bibnamefont {Bouchard}}, \bibinfo {author} {\bibfnamefont
  {I.}~\bibnamefont {Plotnikov}}, \bibinfo {author} {\bibfnamefont
  {N.}~\bibnamefont {Aunai}}, \bibinfo {author} {\bibfnamefont
  {J.}~\bibnamefont {Dargent}}, \bibinfo {author} {\bibfnamefont
  {C.}~\bibnamefont {Riconda}}, \ and\ \bibinfo {author} {\bibfnamefont
  {M.}~\bibnamefont {Grech}},\ }\bibfield  {title} {\enquote {\bibinfo {title}
  {Smilei : A collaborative, open-source, multi-purpose particle-in-cell code
  for plasma simulation},}\ }\href {\doibase
  https://doi.org/10.1016/j.cpc.2017.09.024} {\bibfield  {journal} {\bibinfo
  {journal} {Computer Physics Communications}\ }\textbf {\bibinfo {volume}
  {222}},\ \bibinfo {pages} {351 -- 373} (\bibinfo {year} {2018})}\BibitemShut
  {NoStop}%
\bibitem [{\citenamefont {Sentoku}\ and\ \citenamefont
  {Kemp}(2008)}]{Sentoku:2008aa}%
  \BibitemOpen
  \bibfield  {author} {\bibinfo {author} {\bibfnamefont {Y.}~\bibnamefont
  {Sentoku}}\ and\ \bibinfo {author} {\bibfnamefont {A.~J.}\ \bibnamefont
  {Kemp}},\ }\bibfield  {title} {\enquote {\bibinfo {title} {Numerical methods
  for particle simulations at extreme densities and temperatures: Weighted
  particles, relativistic collisions and reduced currents},}\ }\href {\doibase
  http://dx.doi.org/10.1016/j.jcp.2008.03.043} {\bibfield  {journal} {\bibinfo
  {journal} {Journal of Computational Physics}\ }\textbf {\bibinfo {volume}
  {227}},\ \bibinfo {pages} {6846--6861} (\bibinfo {year} {2008})}\BibitemShut
  {NoStop}%
\bibitem [{\citenamefont {Nanbu}(1997)}]{Nanbu1997}%
  \BibitemOpen
  \bibfield  {author} {\bibinfo {author} {\bibfnamefont {K.}~\bibnamefont
  {Nanbu}},\ }\bibfield  {title} {\enquote {\bibinfo {title} {Theory of
  cumulative small-angle collisions in plasmas},}\ }\href {\doibase
  10.1103/PhysRevE.55.4642} {\bibfield  {journal} {\bibinfo  {journal} {Phys.
  Rev. E}\ }\textbf {\bibinfo {volume} {55}},\ \bibinfo {pages} {4642--4652}
  (\bibinfo {year} {1997})}\BibitemShut {NoStop}%
\bibitem [{\citenamefont {P{\'e}rez}\ \emph {et~al.}(2012)\citenamefont
  {P{\'e}rez}, \citenamefont {Gremillet}, \citenamefont {Decoster},
  \citenamefont {Drouin},\ and\ \citenamefont {Lefebvre}}]{Perez2012}%
  \BibitemOpen
  \bibfield  {author} {\bibinfo {author} {\bibfnamefont {F.}~\bibnamefont
  {P{\'e}rez}}, \bibinfo {author} {\bibfnamefont {L.}~\bibnamefont
  {Gremillet}}, \bibinfo {author} {\bibfnamefont {A.}~\bibnamefont {Decoster}},
  \bibinfo {author} {\bibfnamefont {M.}~\bibnamefont {Drouin}}, \ and\ \bibinfo
  {author} {\bibfnamefont {E.}~\bibnamefont {Lefebvre}},\ }\bibfield  {title}
  {\enquote {\bibinfo {title} {Improved modeling of relativistic collisions and
  collisional ionization in particle-in-cell codes},}\ }\href {\doibase
  10.1063/1.4742167} {\bibfield  {journal} {\bibinfo  {journal} {Physics of
  Plasmas}\ }\textbf {\bibinfo {volume} {19}},\ \bibinfo {pages} {083104}
  (\bibinfo {year} {2012})},\ \Eprint
  {http://arxiv.org/abs/https://doi.org/10.1063/1.4742167}
  {https://doi.org/10.1063/1.4742167} \BibitemShut {NoStop}%
\bibitem [{\citenamefont {Huba}(2013)}]{Huba2013}%
  \BibitemOpen
  \bibfield  {author} {\bibinfo {author} {\bibfnamefont {J.~D.}\ \bibnamefont
  {Huba}},\ }\href {http://wwwppd.nrl.navy.mil/nrlformulary/} {\emph {\bibinfo
  {title} {Plasma Physics}}}\ (\bibinfo  {publisher} {Naval Research
  Laboratory},\ \bibinfo {address} {Washington, DC},\ \bibinfo {year} {2013})\
  pp.\ \bibinfo {pages} {1--71}\BibitemShut {NoStop}%
\bibitem [{\citenamefont {Blandford}\ and\ \citenamefont
  {McKee}(1976)}]{Blandford:1976aa}%
  \BibitemOpen
  \bibfield  {author} {\bibinfo {author} {\bibfnamefont {R.~D.}\ \bibnamefont
  {Blandford}}\ and\ \bibinfo {author} {\bibfnamefont {C.~F.}\ \bibnamefont
  {McKee}},\ }\bibfield  {title} {\enquote {\bibinfo {title} {Fluid dynamics of
  relativistic blast waves},}\ }\href {\doibase 10.1063/1.861619} {\bibfield
  {journal} {\bibinfo  {journal} {Physics of Fluids}\ }\textbf {\bibinfo
  {volume} {19}},\ \bibinfo {pages} {1130--1138} (\bibinfo {year}
  {1976})}\BibitemShut {NoStop}%
\bibitem [{\citenamefont {Stockem}\ \emph {et~al.}(2012)\citenamefont
  {Stockem}, \citenamefont {Fi{\'u}za}, \citenamefont {Fonseca},\ and\
  \citenamefont {Silva}}]{Stockem:2012aa}%
  \BibitemOpen
  \bibfield  {author} {\bibinfo {author} {\bibfnamefont {A.}~\bibnamefont
  {Stockem}}, \bibinfo {author} {\bibfnamefont {F.}~\bibnamefont {Fi{\'u}za}},
  \bibinfo {author} {\bibfnamefont {R.~A.}\ \bibnamefont {Fonseca}}, \ and\
  \bibinfo {author} {\bibfnamefont {L.~O.}\ \bibnamefont {Silva}},\ }\bibfield
  {title} {\enquote {\bibinfo {title} {The impact of kinetic effects on the
  properties of relativistic electron--positron shocks},}\ }\href
  {http://stacks.iop.org/0741-3335/54/i=12/a=125004} {\bibfield  {journal}
  {\bibinfo  {journal} {Plasma Physics and Controlled Fusion}\ }\textbf
  {\bibinfo {volume} {54}},\ \bibinfo {pages} {125004} (\bibinfo {year}
  {2012})}\BibitemShut {NoStop}%
\bibitem [{\citenamefont {Tidman}\ and\ \citenamefont
  {Krall}(1971)}]{Tidman:1971aa}%
  \BibitemOpen
  \bibfield  {author} {\bibinfo {author} {\bibfnamefont {D.~A.}\ \bibnamefont
  {Tidman}}\ and\ \bibinfo {author} {\bibfnamefont {N.~A.}\ \bibnamefont
  {Krall}},\ }\href@noop {} {\emph {\bibinfo {title} {Shock waves in
  collisionaless plasmas}}},\ edited by\ \bibinfo {editor} {\bibfnamefont
  {S.~C.}\ \bibnamefont {Brown}},\ Wiley Series in Plasma Physics\ (\bibinfo
  {publisher} {Wiley-Interscience},\ \bibinfo {year} {1971})\BibitemShut
  {NoStop}%
\bibitem [{\citenamefont {Nakatsutsumi}\ \emph {et~al.}(2018)\citenamefont
  {Nakatsutsumi}, \citenamefont {Sentoku}, \citenamefont {Korzhimanov},
  \citenamefont {Chen}, \citenamefont {Buffechoux}, \citenamefont {Kon},
  \citenamefont {Atherton}, \citenamefont {Audebert}, \citenamefont {Geissel},
  \citenamefont {Hurd}, \citenamefont {Kimmel}, \citenamefont {Rambo},
  \citenamefont {Schollmeier}, \citenamefont {Schwarz}, \citenamefont
  {Starodubtsev}, \citenamefont {Gremillet}, \citenamefont {Kodama},\ and\
  \citenamefont {Fuchs}}]{Nakatsutsumi:2018aa}%
  \BibitemOpen
  \bibfield  {author} {\bibinfo {author} {\bibfnamefont {M.}~\bibnamefont
  {Nakatsutsumi}}, \bibinfo {author} {\bibfnamefont {Y.}~\bibnamefont
  {Sentoku}}, \bibinfo {author} {\bibfnamefont {A.}~\bibnamefont
  {Korzhimanov}}, \bibinfo {author} {\bibfnamefont {S.~N.}\ \bibnamefont
  {Chen}}, \bibinfo {author} {\bibfnamefont {S.}~\bibnamefont {Buffechoux}},
  \bibinfo {author} {\bibfnamefont {A.}~\bibnamefont {Kon}}, \bibinfo {author}
  {\bibfnamefont {B.}~\bibnamefont {Atherton}}, \bibinfo {author}
  {\bibfnamefont {P.}~\bibnamefont {Audebert}}, \bibinfo {author}
  {\bibfnamefont {M.}~\bibnamefont {Geissel}}, \bibinfo {author} {\bibfnamefont
  {L.}~\bibnamefont {Hurd}}, \bibinfo {author} {\bibfnamefont {M.}~\bibnamefont
  {Kimmel}}, \bibinfo {author} {\bibfnamefont {P.}~\bibnamefont {Rambo}},
  \bibinfo {author} {\bibfnamefont {M.}~\bibnamefont {Schollmeier}}, \bibinfo
  {author} {\bibfnamefont {J.}~\bibnamefont {Schwarz}}, \bibinfo {author}
  {\bibfnamefont {M.}~\bibnamefont {Starodubtsev}}, \bibinfo {author}
  {\bibfnamefont {L.}~\bibnamefont {Gremillet}}, \bibinfo {author}
  {\bibfnamefont {R.}~\bibnamefont {Kodama}}, \ and\ \bibinfo {author}
  {\bibfnamefont {J.}~\bibnamefont {Fuchs}},\ }\bibfield  {title} {\enquote
  {\bibinfo {title} {Self-generated surface magnetic fields inhibit
  laser-driven sheath acceleration of high-energy protons},}\ }\href {\doibase
  10.1038/s41467-017-02436-w} {\bibfield  {journal} {\bibinfo  {journal}
  {Nature Communications}\ }\textbf {\bibinfo {volume} {9}},\ \bibinfo {pages}
  {280} (\bibinfo {year} {2018})}\BibitemShut {NoStop}%
\bibitem [{\citenamefont {Bell}\ \emph {et~al.}(1997)\citenamefont {Bell},
  \citenamefont {Davies}, \citenamefont {Guerin},\ and\ \citenamefont
  {Ruhl}}]{Bell:1997aa}%
  \BibitemOpen
  \bibfield  {author} {\bibinfo {author} {\bibfnamefont {A.~R.}\ \bibnamefont
  {Bell}}, \bibinfo {author} {\bibfnamefont {J.~R.}\ \bibnamefont {Davies}},
  \bibinfo {author} {\bibfnamefont {S.}~\bibnamefont {Guerin}}, \ and\ \bibinfo
  {author} {\bibfnamefont {H.}~\bibnamefont {Ruhl}},\ }\bibfield  {title}
  {\enquote {\bibinfo {title} {Fast-electron transport in high-intensity
  short-pulse laser - solid experiments},}\ }\href
  {http://stacks.iop.org/0741-3335/39/i=5/a=001} {\bibfield  {journal}
  {\bibinfo  {journal} {Plasma Physics and Controlled Fusion}\ }\textbf
  {\bibinfo {volume} {39}},\ \bibinfo {pages} {653} (\bibinfo {year}
  {1997})}\BibitemShut {NoStop}%
\bibitem [{\citenamefont {Batani}\ \emph {et~al.}(2002)\citenamefont {Batani},
  \citenamefont {Antonicci}, \citenamefont {Pisani}, \citenamefont {Hall},
  \citenamefont {Scott}, \citenamefont {Amiranoff}, \citenamefont {Koenig},
  \citenamefont {Gremillet}, \citenamefont {Baton}, \citenamefont {Martinolli},
  \citenamefont {Rousseaux},\ and\ \citenamefont {Nazarov}}]{Batani:2002aa}%
  \BibitemOpen
  \bibfield  {author} {\bibinfo {author} {\bibfnamefont {D.}~\bibnamefont
  {Batani}}, \bibinfo {author} {\bibfnamefont {A.}~\bibnamefont {Antonicci}},
  \bibinfo {author} {\bibfnamefont {F.}~\bibnamefont {Pisani}}, \bibinfo
  {author} {\bibfnamefont {T.~A.}\ \bibnamefont {Hall}}, \bibinfo {author}
  {\bibfnamefont {D.}~\bibnamefont {Scott}}, \bibinfo {author} {\bibfnamefont
  {F.}~\bibnamefont {Amiranoff}}, \bibinfo {author} {\bibfnamefont
  {M.}~\bibnamefont {Koenig}}, \bibinfo {author} {\bibfnamefont
  {L.}~\bibnamefont {Gremillet}}, \bibinfo {author} {\bibfnamefont
  {S.}~\bibnamefont {Baton}}, \bibinfo {author} {\bibfnamefont
  {E.}~\bibnamefont {Martinolli}}, \bibinfo {author} {\bibfnamefont
  {C.}~\bibnamefont {Rousseaux}}, \ and\ \bibinfo {author} {\bibfnamefont
  {W.}~\bibnamefont {Nazarov}},\ }\bibfield  {title} {\enquote {\bibinfo
  {title} {Inhibition in the propagation of fast electrons in plastic foams by
  resistive electric fields},}\ }\href {\doibase 10.1103/PhysRevE.65.066409}
  {\bibfield  {journal} {\bibinfo  {journal} {Phys. Rev. E}\ }\textbf {\bibinfo
  {volume} {65}},\ \bibinfo {pages} {066409} (\bibinfo {year}
  {2002})}\BibitemShut {NoStop}%
\bibitem [{\citenamefont {McKenna}\ and\ \citenamefont
  {Quinn}(2013)}]{McKenna:2013aa}%
  \BibitemOpen
  \bibfield  {author} {\bibinfo {author} {\bibfnamefont {P.}~\bibnamefont
  {McKenna}}\ and\ \bibinfo {author} {\bibfnamefont {M.~N.}\ \bibnamefont
  {Quinn}},\ }\enquote {\bibinfo {title} {Laser-plasma interactions and
  applications},}\ \ (\bibinfo  {publisher} {Springer, Heidelberg},\ \bibinfo
  {year} {2013})\ Chap.~\bibinfo {chapter} {5}, pp.\ \bibinfo {pages}
  {91--115}\BibitemShut {NoStop}%
\bibitem [{\citenamefont {Gibbon}(2005)}]{Gibbon:2005aa}%
  \BibitemOpen
  \bibfield  {author} {\bibinfo {author} {\bibfnamefont {P.}~\bibnamefont
  {Gibbon}},\ }\bibfield  {title} {\enquote {\bibinfo {title} {Resistively
  enhanced proton acceleration via high-intensity laser interactions with cold
  foil targets},}\ }\href {\doibase 10.1103/PhysRevE.72.026411} {\bibfield
  {journal} {\bibinfo  {journal} {Phys. Rev. E}\ }\textbf {\bibinfo {volume}
  {72}},\ \bibinfo {pages} {026411} (\bibinfo {year} {2005})}\BibitemShut
  {NoStop}%
\bibitem [{\citenamefont {Robinson}\ \emph {et~al.}(2014)\citenamefont
  {Robinson}, \citenamefont {Strozzi}, \citenamefont {Davies}, \citenamefont
  {Gremillet}, \citenamefont {Honrubia}, \citenamefont {Johzaki}, \citenamefont
  {Kingham}, \citenamefont {Sherlock},\ and\ \citenamefont
  {Solodov}}]{Robinson:2014aa}%
  \BibitemOpen
  \bibfield  {author} {\bibinfo {author} {\bibfnamefont {A.}~\bibnamefont
  {Robinson}}, \bibinfo {author} {\bibfnamefont {D.}~\bibnamefont {Strozzi}},
  \bibinfo {author} {\bibfnamefont {J.}~\bibnamefont {Davies}}, \bibinfo
  {author} {\bibfnamefont {L.}~\bibnamefont {Gremillet}}, \bibinfo {author}
  {\bibfnamefont {J.}~\bibnamefont {Honrubia}}, \bibinfo {author}
  {\bibfnamefont {T.}~\bibnamefont {Johzaki}}, \bibinfo {author} {\bibfnamefont
  {R.}~\bibnamefont {Kingham}}, \bibinfo {author} {\bibfnamefont
  {M.}~\bibnamefont {Sherlock}}, \ and\ \bibinfo {author} {\bibfnamefont
  {A.}~\bibnamefont {Solodov}},\ }\bibfield  {title} {\enquote {\bibinfo
  {title} {Theory of fast electron transport for fast ignition},}\ }\href
  {http://stacks.iop.org/0029-5515/54/i=5/a=054003} {\bibfield  {journal}
  {\bibinfo  {journal} {Nuclear Fusion}\ }\textbf {\bibinfo {volume} {54}},\
  \bibinfo {pages} {054003} (\bibinfo {year} {2014})}\BibitemShut {NoStop}%
\bibitem [{\citenamefont {Pak}\ \emph {et~al.}(2018)\citenamefont {Pak},
  \citenamefont {Kerr}, \citenamefont {Lemos}, \citenamefont {Link},
  \citenamefont {Patel}, \citenamefont {Albert}, \citenamefont {Divol},
  \citenamefont {Pollock}, \citenamefont {Haberberger}, \citenamefont {Froula},
  \citenamefont {Gauthier}, \citenamefont {Glenzer}, \citenamefont {Longman},
  \citenamefont {Manzoor}, \citenamefont {Fedosejevs}, \citenamefont
  {Tochitsky}, \citenamefont {Joshi},\ and\ \citenamefont {Fiuza}}]{Pak2018}%
  \BibitemOpen
  \bibfield  {author} {\bibinfo {author} {\bibfnamefont {A.}~\bibnamefont
  {Pak}}, \bibinfo {author} {\bibfnamefont {S.}~\bibnamefont {Kerr}}, \bibinfo
  {author} {\bibfnamefont {N.}~\bibnamefont {Lemos}}, \bibinfo {author}
  {\bibfnamefont {A.}~\bibnamefont {Link}}, \bibinfo {author} {\bibfnamefont
  {P.}~\bibnamefont {Patel}}, \bibinfo {author} {\bibfnamefont
  {F.}~\bibnamefont {Albert}}, \bibinfo {author} {\bibfnamefont
  {L.}~\bibnamefont {Divol}}, \bibinfo {author} {\bibfnamefont {B.~B.}\
  \bibnamefont {Pollock}}, \bibinfo {author} {\bibfnamefont {D.}~\bibnamefont
  {Haberberger}}, \bibinfo {author} {\bibfnamefont {D.}~\bibnamefont {Froula}},
  \bibinfo {author} {\bibfnamefont {M.}~\bibnamefont {Gauthier}}, \bibinfo
  {author} {\bibfnamefont {S.~H.}\ \bibnamefont {Glenzer}}, \bibinfo {author}
  {\bibfnamefont {A.}~\bibnamefont {Longman}}, \bibinfo {author} {\bibfnamefont
  {L.}~\bibnamefont {Manzoor}}, \bibinfo {author} {\bibfnamefont
  {R.}~\bibnamefont {Fedosejevs}}, \bibinfo {author} {\bibfnamefont
  {S.}~\bibnamefont {Tochitsky}}, \bibinfo {author} {\bibfnamefont
  {C.}~\bibnamefont {Joshi}}, \ and\ \bibinfo {author} {\bibfnamefont
  {F.}~\bibnamefont {Fiuza}},\ }\bibfield  {title} {\enquote {\bibinfo {title}
  {Collisionless shock acceleration of narrow energy spread ion beams from
  mixed species plasmas using $1\text{ }\text{ }\ensuremath{\mu}\mathrm{m}$
  lasers},}\ }\href {\doibase 10.1103/PhysRevAccelBeams.21.103401} {\bibfield
  {journal} {\bibinfo  {journal} {Phys. Rev. Accel. Beams}\ }\textbf {\bibinfo
  {volume} {21}},\ \bibinfo {pages} {103401} (\bibinfo {year}
  {2018})}\BibitemShut {NoStop}%
\bibitem [{\citenamefont {Bret}, \citenamefont {Gremillet},\ and\ \citenamefont
  {B\'enisti}(2010)}]{Bret_2010_exact}%
  \BibitemOpen
  \bibfield  {author} {\bibinfo {author} {\bibfnamefont {A.}~\bibnamefont
  {Bret}}, \bibinfo {author} {\bibfnamefont {L.}~\bibnamefont {Gremillet}}, \
  and\ \bibinfo {author} {\bibfnamefont {D.}~\bibnamefont {B\'enisti}},\
  }\bibfield  {title} {\enquote {\bibinfo {title} {Exact relativistic kinetic
  theory of the full unstable spectrum of an electron-beam--plasma system with
  maxwell-j\"uttner distribution functions},}\ }\href {\doibase
  10.1103/PhysRevE.81.036402} {\bibfield  {journal} {\bibinfo  {journal} {Phys.
  Rev. E}\ }\textbf {\bibinfo {volume} {81}},\ \bibinfo {pages} {036402}
  (\bibinfo {year} {2010})}\BibitemShut {NoStop}%
\bibitem [{\citenamefont {Tzoufras}\ \emph {et~al.}(2006)\citenamefont
  {Tzoufras}, \citenamefont {Ren}, \citenamefont {Tsung}, \citenamefont
  {Tonge}, \citenamefont {Mori}, \citenamefont {Fiore}, \citenamefont
  {Fonseca},\ and\ \citenamefont {Silva}}]{Tzoufras:2006}%
  \BibitemOpen
  \bibfield  {author} {\bibinfo {author} {\bibfnamefont {M.}~\bibnamefont
  {Tzoufras}}, \bibinfo {author} {\bibfnamefont {C.}~\bibnamefont {Ren}},
  \bibinfo {author} {\bibfnamefont {F.~S.}\ \bibnamefont {Tsung}}, \bibinfo
  {author} {\bibfnamefont {J.~W.}\ \bibnamefont {Tonge}}, \bibinfo {author}
  {\bibfnamefont {W.~B.}\ \bibnamefont {Mori}}, \bibinfo {author}
  {\bibfnamefont {M.}~\bibnamefont {Fiore}}, \bibinfo {author} {\bibfnamefont
  {R.~A.}\ \bibnamefont {Fonseca}}, \ and\ \bibinfo {author} {\bibfnamefont
  {L.~O.}\ \bibnamefont {Silva}},\ }\bibfield  {title} {\enquote {\bibinfo
  {title} {Space-charge effects in the current-filamentation or weibel
  instability},}\ }\href {\doibase 10.1103/PhysRevLett.96.105002} {\bibfield
  {journal} {\bibinfo  {journal} {Phys. Rev. Lett.}\ }\textbf {\bibinfo
  {volume} {96}},\ \bibinfo {pages} {105002} (\bibinfo {year}
  {2006})}\BibitemShut {NoStop}%
\bibitem [{\citenamefont {Stockem}\ \emph
  {et~al.}(2014{\natexlab{b}})\citenamefont {Stockem}, \citenamefont
  {Grismayer}, \citenamefont {Fonseca},\ and\ \citenamefont
  {Silva}}]{Stockem:2014aa}%
  \BibitemOpen
  \bibfield  {author} {\bibinfo {author} {\bibfnamefont {A.}~\bibnamefont
  {Stockem}}, \bibinfo {author} {\bibfnamefont {T.}~\bibnamefont {Grismayer}},
  \bibinfo {author} {\bibfnamefont {R.}~\bibnamefont {Fonseca}}, \ and\
  \bibinfo {author} {\bibfnamefont {L.}~\bibnamefont {Silva}},\ }\bibfield
  {title} {\enquote {\bibinfo {title} {Electromagnetic field generation in the
  downstream of electrostatic shocks due to electron trapping},}\ }\href
  {\doibase 10.1103/physrevlett.113.105002} {\bibfield  {journal} {\bibinfo
  {journal} {Physical Review Letters}\ }\textbf {\bibinfo {volume} {113}}
  (\bibinfo {year} {2014}{\natexlab{b}}),\
  10.1103/physrevlett.113.105002}\BibitemShut {NoStop}%
\end{thebibliography}
%

\end{document}